\documentclass[sigconf,balance=false]{acmart}

\usepackage{popets}

\setcopyright{popets}
\copyrightyear{2023}
\acmYear{2023}
\acmVolume{2023}
\acmNumber{X}
\acmDOI{XXXXXXX.XXXXXXX}
\acmISBN{}
\acmConference{Proceedings on Privacy Enhancing Technologies}
\usepackage{lipsum}             
\usepackage{array}
\usepackage{multirow}
\usepackage{ragged2e}
\usepackage{listings}
\usepackage{booktabs,caption}
\usepackage[flushleft]{threeparttable}

\newcommand{\totalAppCount}{160}

\newcommand{\totalCCPAAppCount}{109}

\newcommand{\totalCCPAAppPercentage}{68\%}

\newcommand{\accountHolderAppCount}{91}

\newcommand{\accountHolderAppPercentage}{83\%}

\newcommand{\krippendorffCCPA}{0.81}

\newcommand{\twoOrMoreSARMethodsAppPercentage}{66\%}

\newcommand{\emailMethodCount}{71}
\newcommand{\emailMethodProportion}{0.65}
\newcommand{\emailMethodPercentage}{65\%}

\newcommand{\totalSARPortalMethodCount}{42}

\newcommand{\totalSARPortalMethodPercentage}{39\%}

\newcommand{\oneTrustMethodCount}{15}
\newcommand{\oneTrustMethodProportion}{0.14}

\newcommand{\ownSARPortalMethodCount}{27}
\newcommand{\ownSARPortalMethodProportion}{0.25}

\newcommand{\phoneMethodCount}{27}
\newcommand{\phoneMethodProportion}{0.25}
\newcommand{\phoneMethodPercentage}{25\%}

\newcommand{\mailMethodCount}{21}
\newcommand{\mailMethodProportion}{0.19}
\newcommand{\mailMethodPercentage}{19\%}

\newcommand{\customerSupportMethodCount}{21}
\newcommand{\customerSupportMethodProportion}{0.19}
\newcommand{\customerSupportMethodPercentage}{19\%}

\newcommand{\accountPrivacyMethodCount}{11}
\newcommand{\accountPrivacyMethodProportion}{0.10}

\newcommand{\appPrivacyMethodCount}{5}
\newcommand{\appPrivacyMethodProportion}{0.05}

\newcommand{\settingsMethodPercentage}{15\%}

\newcommand{\appFeedbackMethodCount}{3}
\newcommand{\appFeedbackMethodProportion}{0.03}

\newcommand{\googleFormMethodCount}{2}
\newcommand{\googleFormMethodProportion}{0.02}
\newcommand{\googleFormMethodPercentage}{2\%}

\newcommand{\alternativeRequestCounts}{16}

\newcommand{\emailRequestCounts}{52}

\newcommand{\DSARRequestPercentage}{34\%}

\newcommand{\SettingsRequestPercentage}{13\%}

\newcommand{\CustomerSupportRequestCounts}{6}

\newcommand{\FreeFormRequestCounts}{58}

\newcommand{\PreparedFormRequestCounts}{51}

\newcommand{\FreeFormConfirmedReceiptCounts}{40}

\newcommand{\FreeFormConfirmedReceiptPercentage}{69\%}

\newcommand{\FreeFormNotConfirmedReceiptCounts}{18}

\newcommand{\FreeFormNotConfirmedReceiptPercentage}{31\%}

\newcommand{\NotEmailMethodVerifiedEmailYesCount}{32}

\newcommand{\VerifiedAccountYesCount}{35}

\newcommand{\VerificationEmailCount}{36}

\newcommand{\VerificationNameCount}{26}

\newcommand{\VerificationStateCount}{21}

\newcommand{\VerificationAppSpecificInfoCount}{18}

\newcommand{\VerificationCountryCount}{15}

\newcommand{\VerificationUsernameCount}{9}

\newcommand{\VerificationPhoneNumberCount}{7}

\newcommand{\VerificationPostalAddressCount}{6}

\newcommand{\VerificationDeviceUserIDCount}{5}

\newcommand{\VerificationAAIDCount}{5}

\newcommand{\VerificationCertificationPerjuryCount}{5}

\newcommand{\VerificationAffidavitCount}{4}

\newcommand{\VerificationGovernmentIDCount}{3}

\newcommand{\VerificationIPCount}{2}

\newcommand{\VerificationPhoneNumberOwnershipCount}{3}

\newcommand{\VerificationDOBCount}{2}

\newcommand{\VerificationIDMeCount}{1}

\newcommand{\VerificationCallCount}{1}

\newcommand{\CAResidencyYesCount}{7}

\newcommand{\CAResidencyNoCount}{102}

\newcommand{\CAResidencyNoPercentage}{94\%}

\newcommand{\ResponseWithDataCount}{69}

\newcommand{\ResponseWithDataPercentage}{63\%}

\newcommand{\NoResponseCount}{21}

\newcommand{\NoResponsePercentage}{19\%}

\newcommand{\ResponseWithoutDataCount}{8}

\newcommand{\ResponseWithoutDataPercentage}{7\%}

\newcommand{\NoResponseCannotVerifyCount}{5}

\newcommand{\NoResponseCannotVerifyPercentage}{5\%}

\newcommand{\ResponseCannotComplyCount}{3}

\newcommand{\ResponseCannotComplyPercentage}{3\%}

\newcommand{\ResponseSettingsCount}{3}

\newcommand{\ResponseSettingsPercentage}{3\%}

\newcommand{\ResponseNoCount}{29}

\newcommand{\ResponseYesCount}{80}

\newcommand{\HumanResponseCount}{32}

\newcommand{\HumanResponsePercentage}{40\%}

\newcommand{\AutomatedResponseCount}{48}

\newcommand{\AutomatedResponsePercentage}{60\%}

\newcommand{\ActionsMedian}{1}
\newcommand{\ActionsMean}{1.8}
\newcommand{\ActionsSD}{0.78}

\newcommand{\ActionsMax}{4}

\newcommand{\DurationDaysMedian}{7}
\newcommand{\DurationDaysMean}{14.86}
\newcommand{\DurationDaysSD}{18.86}

\newcommand{\DurationDaysMax}{76}

\newcommand{\specificPIIYesCount}{68}

\newcommand{\collectedCategoriesPIIYesCount}{24}

\newcommand{\collectedCategoriesPIIYesPercentage}{35\%}

\newcommand{\disclosedCategoriesPIIYesCount}{18}

\newcommand{\disclosedCategoriesPIIYesPercentage}{26\%}

\newcommand{\partyCategoriesYesCount}{25}

\newcommand{\partyCategoriesYesPercentage}{36\%}

\newcommand{\purposeYesCount}{30}

\newcommand{\purposeYesPercentage}{43\%}

\newcommand{\sourcesYesCount}{23}

\newcommand{\sourcesYesPercentage}{33\%}

\newcommand{\emailResponseCount}{23}

\newcommand{\emailResponsePercentage}{33\%}

\newcommand{\dsarResponseCount}{19}

\newcommand{\dsarResponsePercentage}{28\%}

\newcommand{\settingsResponseCount}{12}

\newcommand{\settingsResponsePercentage}{17\%}

\newcommand{\remainingResponseCount}{15}

\newcommand{\accessExpiresYesCount}{43}

\newcommand{\accessPasswordYesCount}{9}

\newcommand{\accessEmailYesCount}{16}

\newcommand{\accessAccountYesCount}{26}

\newcommand{\accessOtherYesCount}{3}

\newcommand{\accessConfidentialYesCount}{2}

\newcommand{\dataFilesCount}{62}

\newcommand{\dataCSVCount}{27}

\newcommand{\dataJSONCount}{18}

\newcommand{\dataPDFCount}{12}

\newcommand{\dataXLSXCount}{11}

\newcommand{\dataTXTCount}{9}

\newcommand{\dataOneOptionCount}{56}

\newcommand{\dataTwoOptionCount}{6}

\newcommand{\disclosedProvidedSpecificCount}{9}

\newcommand{\undisclosedProvidedSpecificIdentifiersCount}{55}

\newcommand{\undisclosedProvidedSpecificGeolocationCount}{21}

\newcommand{\undisclosedProvidedSpecificSensoryCount}{18}

\newcommand{\undisclosedProvidedSpecificRecordsCount}{16}

\newcommand{\undisclosedProvidedSpecificEmploymentCount}{4}

\newcommand{\undisclosedProvidedSpecificProtectedCount}{3}

\newcommand{\undisclosedProvidedSpecificDeviceIdentifiersCount}{51}

\newcommand{\undisclosedProvidedSpecificAppIdentifiersCount}{28}

\newcommand{\undisclosedProvidedSpecificCoarseLatLonCount}{4}

\newcommand{\undisclosedProvidedSpecificPreciseLatLonCount}{12}

\newcommand{\undisclosedProvidedSpecificZIPCount}{8}

\newcommand{\undisclosedProvidedSpecificCityCount}{12}

\newcommand{\undisclosedProvidedSpecificResidenceRecordsCount}{10}

\newcommand{\undisclosedProvidedSpecificPhoneCount}{5}

\newcommand{\undisclosedProvidedSpecificContactRecordsCount}{5}
\newcommand{\undisclosedClaimedNotCollectedCount}{7}

\newcommand{\undisclosedClaimedNotCollectedIdentifiersCount}{7}

\newcommand{\undisclosedClaimedNotCollectedGeolocationCount}{3}

\newcommand{\undisclosedClaimedNotCollectedSensoryCount}{3}

\newcommand{\undisclosedClaimedNotCollectedAAIDCount}{7}

\newcommand{\undisclosedClaimedNotCollectedIPCount}{5}

\newcommand{\undisclosedClaimedNotCollectedCoarseGeoCount}{3}

\newcommand{\undisclosedClaimedNotCollectedSpecificSensorCount}{3}

\newcommand{\undisclosedAaidCounts}{49}
\newcommand{\undisclosedAaidTLS}{44}
\newcommand{\undisclosedAaidFirstParty}{32}
\newcommand{\undisclosedAaidThirdParty}{43}

\newcommand{\undisclosedAaidPercentageTLS}{89.8\%}

\newcommand{\undisclosedAaidPercentageFirstParty}{65.3\%}

\newcommand{\undisclosedAaidPercentageThirdParty}{87.8\%}

\newcommand{\undisclosedFingerprintidCounts}{25}
\newcommand{\undisclosedFingerprintidTLS}{23}
\newcommand{\undisclosedFingerprintidFirstParty}{2}
\newcommand{\undisclosedFingerprintidThirdParty}{25}

\newcommand{\undisclosedFingerprintidPercentageTLS}{92\%}

\newcommand{\undisclosedFingerprintidPercentageFirstParty}{8\%}

\newcommand{\undisclosedFingerprintidPercentageThirdParty}{100\%}

\newcommand{\undisclosedIpaddressCounts}{23}
\newcommand{\undisclosedIpaddressTLS}{21}
\newcommand{\undisclosedIpaddressFirstParty}{9}
\newcommand{\undisclosedIpaddressThirdParty}{20}

\newcommand{\undisclosedIpaddressPercentageTLS}{91.3\%}

\newcommand{\undisclosedIpaddressPercentageFirstParty}{39.1\%}

\newcommand{\undisclosedIpaddressPercentageThirdParty}{87\%}

\newcommand{\undisclosedSensordataCounts}{22}
\newcommand{\undisclosedSensordataTLS}{22}
\newcommand{\undisclosedSensordataFirstParty}{1}
\newcommand{\undisclosedSensordataThirdParty}{22}

\newcommand{\undisclosedSensordataPercentageTLS}{100\%}

\newcommand{\undisclosedSensordataPercentageFirstParty}{4.5\%}

\newcommand{\undisclosedSensordataPercentageThirdParty}{100\%}

\newcommand{\undisclosedAndroididCounts}{20}
\newcommand{\undisclosedAndroididTLS}{17}
\newcommand{\undisclosedAndroididFirstParty}{10}
\newcommand{\undisclosedAndroididThirdParty}{18}

\newcommand{\undisclosedAndroididPercentageTLS}{85\%}

\newcommand{\undisclosedAndroididPercentageFirstParty}{50\%}

\newcommand{\undisclosedAndroididPercentageThirdParty}{90\%}

\newcommand{\undisclosedIdentityidCounts}{16}
\newcommand{\undisclosedIdentityidTLS}{15}
\newcommand{\undisclosedIdentityidFirstParty}{1}
\newcommand{\undisclosedIdentityidThirdParty}{16}

\newcommand{\undisclosedIdentityidPercentageTLS}{93.8\%}

\newcommand{\undisclosedIdentityidPercentageFirstParty}{6.2\%}

\newcommand{\undisclosedIdentityidPercentageThirdParty}{100\%}

\newcommand{\undisclosedCurrentcityCounts}{15}
\newcommand{\undisclosedCurrentcityTLS}{14}
\newcommand{\undisclosedCurrentcityFirstParty}{8}
\newcommand{\undisclosedCurrentcityThirdParty}{11}

\newcommand{\undisclosedCurrentcityPercentageTLS}{93.3\%}

\newcommand{\undisclosedCurrentcityPercentageFirstParty}{53.3\%}

\newcommand{\undisclosedCurrentcityPercentageThirdParty}{73.3\%}

\newcommand{\undisclosedPreciselongitudelatitudeCounts}{13}
\newcommand{\undisclosedPreciselongitudelatitudeTLS}{10}
\newcommand{\undisclosedPreciselongitudelatitudeFirstParty}{10}
\newcommand{\undisclosedPreciselongitudelatitudeThirdParty}{10}

\newcommand{\undisclosedPreciselongitudelatitudePercentageTLS}{76.9\%}

\newcommand{\undisclosedPreciselongitudelatitudePercentageFirstParty}{76.9\%}

\newcommand{\undisclosedPreciselongitudelatitudePercentageThirdParty}{76.9\%}

\newcommand{\undisclosedAppfingerprintidCounts}{10}
\newcommand{\undisclosedAppfingerprintidTLS}{7}
\newcommand{\undisclosedAppfingerprintidFirstParty}{8}
\newcommand{\undisclosedAppfingerprintidThirdParty}{8}

\newcommand{\undisclosedAppfingerprintidPercentageTLS}{70\%}

\newcommand{\undisclosedAppfingerprintidPercentageFirstParty}{80\%}

\newcommand{\undisclosedAppfingerprintidPercentageThirdParty}{80\%}

\newcommand{\undisclosedCurrentzipCounts}{9}
\newcommand{\undisclosedCurrentzipTLS}{7}
\newcommand{\undisclosedCurrentzipFirstParty}{5}
\newcommand{\undisclosedCurrentzipThirdParty}{7}

\newcommand{\undisclosedCurrentzipPercentageTLS}{77.8\%}

\newcommand{\undisclosedCurrentzipPercentageFirstParty}{55.6\%}

\newcommand{\undisclosedCurrentzipPercentageThirdParty}{77.8\%}

\newcommand{\undisclosedRoutermacCounts}{8}
\newcommand{\undisclosedRoutermacTLS}{8}
\newcommand{\undisclosedRoutermacFirstParty}{2}
\newcommand{\undisclosedRoutermacThirdParty}{6}

\newcommand{\undisclosedRoutermacPercentageTLS}{100\%}

\newcommand{\undisclosedRoutermacPercentageFirstParty}{25\%}

\newcommand{\undisclosedRoutermacPercentageThirdParty}{75\%}

\newcommand{\undisclosedRouterssidCounts}{8}
\newcommand{\undisclosedRouterssidTLS}{7}
\newcommand{\undisclosedRouterssidFirstParty}{3}
\newcommand{\undisclosedRouterssidThirdParty}{6}

\newcommand{\undisclosedRouterssidPercentageTLS}{87.5\%}

\newcommand{\undisclosedRouterssidPercentageFirstParty}{37.5\%}

\newcommand{\undisclosedRouterssidPercentageThirdParty}{75\%}

\newcommand{\undisclosedPostalzipCounts}{6}
\newcommand{\undisclosedPostalzipTLS}{5}
\newcommand{\undisclosedPostalzipFirstParty}{5}
\newcommand{\undisclosedPostalzipThirdParty}{3}

\newcommand{\undisclosedPostalzipPercentageTLS}{83.3\%}

\newcommand{\undisclosedPostalzipPercentageFirstParty}{83.3\%}

\newcommand{\undisclosedPostalzipPercentageThirdParty}{50\%}

\newcommand{\undisclosedPostalcityCounts}{5}
\newcommand{\undisclosedPostalcityTLS}{4}
\newcommand{\undisclosedPostalcityFirstParty}{4}
\newcommand{\undisclosedPostalcityThirdParty}{3}

\newcommand{\undisclosedPostalcityPercentageTLS}{80\%}

\newcommand{\undisclosedPostalcityPercentageFirstParty}{80\%}

\newcommand{\undisclosedPostalcityPercentageThirdParty}{60\%}

\newcommand{\undisclosedCoarselongitudelatitudeCounts}{5}
\newcommand{\undisclosedCoarselongitudelatitudeTLS}{5}
\newcommand{\undisclosedCoarselongitudelatitudeFirstParty}{3}
\newcommand{\undisclosedCoarselongitudelatitudeThirdParty}{2}

\newcommand{\undisclosedCoarselongitudelatitudePercentageTLS}{100\%}

\newcommand{\undisclosedCoarselongitudelatitudePercentageFirstParty}{60\%}

\newcommand{\undisclosedCoarselongitudelatitudePercentageThirdParty}{40\%}

\newcommand{\undisclosedContactnumberCounts}{5}
\newcommand{\undisclosedContactnumberTLS}{5}
\newcommand{\undisclosedContactnumberFirstParty}{5}
\newcommand{\undisclosedContactnumberThirdParty}{0}

\newcommand{\undisclosedContactnumberPercentageTLS}{100\%}

\newcommand{\undisclosedContactnumberPercentageFirstParty}{100\%}

\newcommand{\undisclosedContactnumberPercentageThirdParty}{0\%}

\newcommand{\undisclosedTelephonenumberCounts}{5}
\newcommand{\undisclosedTelephonenumberTLS}{4}
\newcommand{\undisclosedTelephonenumberFirstParty}{5}
\newcommand{\undisclosedTelephonenumberThirdParty}{1}

\newcommand{\undisclosedTelephonenumberPercentageTLS}{80\%}

\newcommand{\undisclosedTelephonenumberPercentageFirstParty}{100\%}

\newcommand{\undisclosedTelephonenumberPercentageThirdParty}{20\%}

\newcommand{\undisclosedDateofbirthCounts}{5}
\newcommand{\undisclosedDateofbirthTLS}{5}
\newcommand{\undisclosedDateofbirthFirstParty}{4}
\newcommand{\undisclosedDateofbirthThirdParty}{1}

\newcommand{\undisclosedDateofbirthPercentageTLS}{100\%}

\newcommand{\undisclosedDateofbirthPercentageFirstParty}{80\%}

\newcommand{\undisclosedDateofbirthPercentageThirdParty}{20\%}

\newcommand{\undisclosedCurrentcountyCounts}{3}
\newcommand{\undisclosedCurrentcountyTLS}{3}
\newcommand{\undisclosedCurrentcountyFirstParty}{3}
\newcommand{\undisclosedCurrentcountyThirdParty}{1}

\newcommand{\undisclosedCurrentcountyPercentageTLS}{100\%}

\newcommand{\undisclosedCurrentcountyPercentageFirstParty}{100\%}

\newcommand{\undisclosedCurrentcountyPercentageThirdParty}{33.3\%}

\newcommand{\undisclosedCompanyCounts}{3}
\newcommand{\undisclosedCompanyTLS}{3}
\newcommand{\undisclosedCompanyFirstParty}{3}
\newcommand{\undisclosedCompanyThirdParty}{0}

\newcommand{\undisclosedCompanyPercentageTLS}{100\%}

\newcommand{\undisclosedCompanyPercentageFirstParty}{100\%}

\newcommand{\undisclosedCompanyPercentageThirdParty}{0\%}

\newcommand{\undisclosedPostalstreetCounts}{3}
\newcommand{\undisclosedPostalstreetTLS}{3}
\newcommand{\undisclosedPostalstreetFirstParty}{1}
\newcommand{\undisclosedPostalstreetThirdParty}{2}

\newcommand{\undisclosedPostalstreetPercentageTLS}{100\%}

\newcommand{\undisclosedPostalstreetPercentageFirstParty}{33.3\%}

\newcommand{\undisclosedPostalstreetPercentageThirdParty}{66.7\%}

\newcommand{\undisclosedDeviceusernameCounts}{3}
\newcommand{\undisclosedDeviceusernameTLS}{3}
\newcommand{\undisclosedDeviceusernameFirstParty}{0}
\newcommand{\undisclosedDeviceusernameThirdParty}{3}

\newcommand{\undisclosedDeviceusernamePercentageTLS}{100\%}

\newcommand{\undisclosedDeviceusernamePercentageFirstParty}{0\%}

\newcommand{\undisclosedDeviceusernamePercentageThirdParty}{100\%}

\newcommand{\undisclosedImeiCounts}{3}
\newcommand{\undisclosedImeiTLS}{3}
\newcommand{\undisclosedImeiFirstParty}{2}
\newcommand{\undisclosedImeiThirdParty}{1}

\newcommand{\undisclosedImeiPercentageTLS}{100\%}

\newcommand{\undisclosedImeiPercentageFirstParty}{66.7\%}

\newcommand{\undisclosedImeiPercentageThirdParty}{33.3\%}

\newcommand{\undisclosedJobCounts}{2}
\newcommand{\undisclosedJobTLS}{1}
\newcommand{\undisclosedJobFirstParty}{2}
\newcommand{\undisclosedJobThirdParty}{1}

\newcommand{\undisclosedJobPercentageTLS}{50\%}

\newcommand{\undisclosedJobPercentageFirstParty}{100\%}

\newcommand{\undisclosedJobPercentageThirdParty}{50\%}

\newcommand{\undisclosedContactnameCounts}{2}
\newcommand{\undisclosedContactnameTLS}{2}
\newcommand{\undisclosedContactnameFirstParty}{2}
\newcommand{\undisclosedContactnameThirdParty}{0}

\newcommand{\undisclosedContactnamePercentageTLS}{100\%}

\newcommand{\undisclosedContactnamePercentageFirstParty}{100\%}

\newcommand{\undisclosedContactnamePercentageThirdParty}{0\%}

\newcommand{\undisclosedImsiCounts}{2}
\newcommand{\undisclosedImsiTLS}{2}
\newcommand{\undisclosedImsiFirstParty}{1}
\newcommand{\undisclosedImsiThirdParty}{1}

\newcommand{\undisclosedImsiPercentageTLS}{100\%}

\newcommand{\undisclosedImsiPercentageFirstParty}{50\%}

\newcommand{\undisclosedImsiPercentageThirdParty}{50\%}

\newcommand{\undisclosedSimidCounts}{2}
\newcommand{\undisclosedSimidTLS}{2}
\newcommand{\undisclosedSimidFirstParty}{1}
\newcommand{\undisclosedSimidThirdParty}{1}

\newcommand{\undisclosedSimidPercentageTLS}{100\%}

\newcommand{\undisclosedSimidPercentageFirstParty}{50\%}

\newcommand{\undisclosedSimidPercentageThirdParty}{50\%}

\newcommand{\undisclosedPostalcountyCounts}{2}
\newcommand{\undisclosedPostalcountyTLS}{2}
\newcommand{\undisclosedPostalcountyFirstParty}{2}
\newcommand{\undisclosedPostalcountyThirdParty}{1}

\newcommand{\undisclosedPostalcountyPercentageTLS}{100\%}

\newcommand{\undisclosedPostalcountyPercentageFirstParty}{100\%}

\newcommand{\undisclosedPostalcountyPercentageThirdParty}{50\%}

\newcommand{\undisclosedWifimacCounts}{1}
\newcommand{\undisclosedWifimacTLS}{1}
\newcommand{\undisclosedWifimacFirstParty}{1}
\newcommand{\undisclosedWifimacThirdParty}{0}

\newcommand{\undisclosedWifimacPercentageTLS}{100\%}

\newcommand{\undisclosedWifimacPercentageFirstParty}{100\%}

\newcommand{\undisclosedWifimacPercentageThirdParty}{0\%}

\newcommand{\undisclosedGenderCounts}{1}
\newcommand{\undisclosedGenderTLS}{1}
\newcommand{\undisclosedGenderFirstParty}{1}
\newcommand{\undisclosedGenderThirdParty}{1}

\newcommand{\undisclosedGenderPercentageTLS}{100\%}

\newcommand{\undisclosedGenderPercentageFirstParty}{100\%}

\newcommand{\undisclosedGenderPercentageThirdParty}{100\%}

\newcommand{\undisclosedHardwareidCounts}{1}
\newcommand{\undisclosedHardwareidTLS}{1}
\newcommand{\undisclosedHardwareidFirstParty}{0}
\newcommand{\undisclosedHardwareidThirdParty}{1}

\newcommand{\undisclosedHardwareidPercentageTLS}{100\%}

\newcommand{\undisclosedHardwareidPercentageFirstParty}{0\%}

\newcommand{\undisclosedHardwareidPercentageThirdParty}{100\%}

\newcommand{\undisclosedCollegeCounts}{1}
\newcommand{\undisclosedCollegeTLS}{1}
\newcommand{\undisclosedCollegeFirstParty}{1}
\newcommand{\undisclosedCollegeThirdParty}{0}

\newcommand{\undisclosedCollegePercentageTLS}{100\%}

\newcommand{\undisclosedCollegePercentageFirstParty}{100\%}

\newcommand{\undisclosedCollegePercentageThirdParty}{0\%}

\begin{document}

\title{Lessons in VCR Repair: Compliance of Android App Developers with the California Consumer Privacy Act (CCPA)}

 \author{Nikita Samarin,$^{1,2}$ Shayna Kothari,$^1$ Zaina Siyed,$^1$ Oscar Bjorkman,$^1$ Reena Yuan,$^1$ Primal Wijesekera,$^{1,2}$ Noura Alomar,$^1$ Jordan Fischer,$^{1,3}$ Chris Hoofnagle,$^{1}$ and Serge Egelman$^{1,2}$}
 \email{{nsamarin, shayna.kothari, zainasiyed, oscarb, reenayuan, primal, nnalomar, jordan.fischer, choofnagle, egelman}@berkeley.edu}
 \affiliation{%
   \institution{$^1$UC Berkeley, $^2$ICSI, $^3$Drexel Kline School of Law}
   \city{Berkeley}
   \state{CA}
   \country{USA}
 }

\renewcommand{\shortauthors}{N. Samarin et al.}

\begin{abstract}
The California Consumer Privacy Act (CCPA) provides California residents with a range of enhanced privacy protections and rights. Our research investigated the extent to which Android app developers comply with the provisions of the CCPA that require them to provide consumers with accurate privacy notices and respond to ``verifiable consumer requests'' (VCRs) by disclosing personal information that they have collected, used, or shared about consumers for a business or commercial purpose. We compared the actual network traffic of 109 apps that we believe must comply with the CCPA to the data that apps state they collect in their privacy policies and the data contained in responses to ``right to know'' requests that we submitted to the app's developers. Of the 69 app developers who substantively replied to our requests, all but one provided specific pieces of personal data (as opposed to only categorical information). However, a significant percentage of apps collected information that was not disclosed, including \textit{identifiers} (55 apps, 80\%), \textit{geolocation data} (21 apps, 30\%), and \textit{sensory data} (18 apps, 26\%) among other categories. We discuss improvements to the CCPA that could help app developers comply with ``right to know'' requests and other related regulations.
\end{abstract}
 

\maketitle

\section{Introduction}
\label{sec:intro}

On January 1, 2020, the California Consumer Privacy Act (CCPA) went into effect. Modeled after the European Union's General Data Protection Regulation (GDPR), the CCPA is designed to increase the control of California consumers over their personal information and offer stronger privacy protections than those available to data subjects in the rest of the United States. Among other provisions, the CCPA requires certain companies operating in California to disclose their data collection and sharing practices and respond to consumers' requests to access their personal information held by the company. This ``right to know'' allows individuals to obtain information that belongs to them and confirm that businesses comply with the data practices stated in their privacy notices. 

The required notice of data practices and the right to know what personal information was collected by a business embody two crucial principles of data protection: individual participation and openness~\cite{organization1980oecd}. Businesses comply with these principles by posting privacy policies and responding to ``subject access requests'' (SARs) from consumers (known as ``verifiable consumer requests'' or ``VCRs'' under the CCPA). Although these principles appear in other privacy frameworks, regulations such as the GDPR and the CCPA define a stricter set of requirements and impose heavier penalties for non-compliance than previous data privacy regimes. For instance, the CCPA prescribes what businesses need to include in their privacy notices and how they should respond to VCRs.

When implemented correctly, the ``right to know'' can greatly benefit consumers. First, accurate information about data collection and sharing practices is necessary to allow consumers to make informed decisions about whether and what information to disclose to the business or whether to seek alternatives, if necessary. Second, the ability to request data pertaining to oneself allows consumers to amend inaccurate information held by the business (the right to rectification) or transmit information to another business of their choosing (the right to data portability). Awareness of the information held by the business can also prompt consumers to request data relating to them be deleted (the right to erasure)~\cite{kemp2020concealed} and lead to the adoption of other privacy-enhancing technologies (PETs). As such, the right to know and other privacy rights enabled by it serve to advance consumers' informational self-determination and increase their bargaining power in digital environments.

Unfortunately, scholarship has already identified shortcomings of other privacy rights granted by the CCPA. For instance, Consumer Reports found that consumers struggled to opt out of the sale of their personal information and were at least \textit{``somewhat dissatisfied''} with the processes they had to go through 52\% of the time~\cite{mahoney2020california}. More recently, Nortwick and Wilson~\cite{van2022setting} found that many websites required to comply with CCPA either failed to provide users with options to request not selling their data to third parties or provided options that suffered from major usability issues. Other studies have also found issues with similar privacy laws enacted earlier, most notably the GDPR in Europe, including evidence of non-compliance by app developers~\cite{zimmeck2019maps} and personal information leakage by abusing the right of access~\cite{di2019personal}. These shortcomings have to be addressed to ensure that the regulations' stated goal of furthering privacy protections for consumers is adequately fulfilled. 

Although prior studies have focused on the impacts of the CCPA and the GDPR~\cite{veys2021pursuing, di2019personal, di2022revisiting, kroger2020app}, we were unable to find any empirical studies measuring the compliance of businesses with the ``right to know'' requirements set by the CCPA, specifically in the context of mobile applications (``apps''). We thus pose the following research question: \textbf{To what extent do Android app developers comply with the provisions of the CCPA that require them to maintain accurate privacy notices and respond to consumers' access requests by disclosing personal information that they have collected about them?} We focus on mobile apps in large part because they present inherent and unique privacy risks, as the devices they are installed on accompany their users throughout their everyday lives and provide access to a wide range of sensitive information, including geolocation, health, and biometric data.

We examined the data practices of \totalAppCount~top-ranked Android mobile app developers from the U.S. Google Play Store, who we expected to meet the definition of a ``business'' regulated under the CCPA and, thus, be required to comply with its provisions. Due to ethical concerns, we focused only on the subset that publicly posted information indicating they would be responsive to users' CCPA requests. We then submitted VCRs to these \totalCCPAAppCount~companies by following the CCPA-specific disclosures in their privacy policies, and compared their responses with the actual data practices that we identified through static and dynamic analysis of their mobile apps. We found that at least 39\% of the apps shared device-specific identifiers and at least 26\% shared geolocation information with third parties without disclosing it in response to our requests. Furthermore, of the \ResponseWithDataCount~app developers who substantially responded to our requests, all but one disclosed the specific pieces of collected personal information, but only \partyCategoriesYesPercentage~included the CCPA-required categories of third-party data recipients in their responses.

The results of our work hold several important policy implications. We argue that regulators---and, in particular, the newly-formed California Privacy Protection Agency (CPPA)---should issue more guidance for developers to help them better comply with the CCPA and its latest amendment, the California Privacy Rights Act (CPRA). Such guidance should include examples of personal information that can be collected from consumers' mobile devices and emphasize the legal obligations for developers who meet the definition of a ``business'' regulated by CCPA. One such obligation is to provide accurate responses to consumers' VCRs; regulators should remind developers that they have to provide all of the requested information, including the categories of personal information and third parties, and ensure that the provided categories are specific to the consumer in question.

\section{Background and Related Work}
\label{sec:background}
We provide an overview of the CCPA, including information about the required notices and disclosures to consumers. We then highlight prior work that investigated the accuracy of disclosures made in privacy policies, the efficacy of subject access request mechanisms, and the potential privacy violations that exist in online systems, including mobile apps and web-based systems.

\subsection{Overview of CCPA's Requirements}
The California Consumer Privacy Act (CCPA) is a state statute that was signed into law in June 2018, becoming effective on January 1, 2020 and enforceable on July 1 of the same year~\cite{ccpa}. The CCPA secures a number of privacy rights for California consumers and imposes new obligations on companies operating in California. In contrast to the EU's General Data Protection Regulation (GDPR), the CCPA only applies to for-profit businesses that do business in California and meet any of the following conditions~\cite{ccpa, gdpr-ref}:
\begin{itemize}
    \item Have a gross annual revenue of over \$25 million;
    \item Buy, receive, or sell the personal information of 50,000 or more California residents, households, or devices; or
    \item Derive 50\% or more of their annual revenue from selling California residents’ personal information.
\end{itemize}

Importantly, the CCPA grants consumers the \textit{right to be notified} about the data collection and sharing practices of a business and, after such collection has taken place, the \textit{right to know} the personal information that the business has pertaining to them. 

\noindent\textbf{Notices to Consumers.} The CCPA requires businesses to provide consumers with a \textit{privacy policy} and a \textit{notice at collection}. The purpose of the privacy policy is ``to provide consumers with a comprehensive description of a business's online and offline practices regarding the collection, use, disclosure, and sale of personal information and of the rights of consumers regarding their personal information''~\cite{ccpa}. The CCPA regulations require that the privacy policy is ``posted online through a conspicuous link using the word `privacy' [...] on the download or landing page of a mobile application'' and include the following information~\cite{ccpa}:
\begin{itemize}
    \item Explanation that a consumer has the right to request that the business disclose what personal information it collects, uses, discloses, and sells;
    \item Instructions for submitting a verifiable consumer request;
    \item Description of the process for verifying the consumer request, including information the consumer must provide;
    \item Categories of personal information the business has collected about consumers in the preceding 12 months;
    \item Categories of personal information, if any, that the business has disclosed or sold in the preceding 12 months and, for each category, the categories of third parties with whom the information was shared;
    \item Categories of sources from which the personal information is collected; and
    \item Business or commercial purpose for collecting or selling personal information.
\end{itemize}

In addition to the privacy policy, the CCPA requires businesses to provide consumers ``with timely notice, at or before the point of collection, about the categories of personal information to be collected from them and the [collection] purposes'' in the form of a notice at collection~\cite{ccpa}. Although businesses might choose to maintain a separate notice at collection, they can also provide a link to the section of the privacy policy containing the required information, as long as the company presents the link at or before the collection of personal information~\cite{ccpa}.

\noindent\textbf{Verifiable Consumer Requests.} The CCPA grants another fundamental privacy right to California consumers, namely, the right to know the personal information that a business has collected pertaining to them. Consumers can exercise this right by submitting a ``verifiable consumer request'' (VCR). The CCPA requires businesses to provide two or more designated methods for submitting VCRs. Furthermore, businesses have 10 days to confirm the receipt of the VCR and 45 days to complete the request, either by providing the requested data or denying it. The CCPA allows businesses to extend the timeline by up to an additional 45 days, provided they inform the requester of the extension and its reasons.

As part of the VCR, consumers can request the same types of information that is required to be in a privacy policy (see list above). However, unlike the general data practices described in the privacy policy, the response to the VCR has to be specific to the consumer making the request. Crucially, in addition to the aforementioned information, a consumer can also request that the business disclose \textit{specific} pieces of personal information that it has collected about the consumer. Unlike the GDPR~\cite{Perry2023}, the CCPA does not require companies to disclose specific names of third parties with whom they share the consumer's personal information.

The CCPA regulations describe the steps that businesses must take to verify the identity of the consumer submitting the VCR. Such verification is crucial to ensure that the company does not disclose a consumer's personal information to an unauthorized party. Simultaneously, businesses need to carefully consider the type and sensitivity of personal information to ensure that their verification procedures do not prevent consumers from successfully exercising their privacy rights. Furthermore, a business should avoid collecting additional personal information solely for the purposes of identity verification (unless absolutely necessary), it cannot impose fees for verification, and should implement reasonable security measures to prevent unauthorized disclosure of consumers' personal information. If a business maintains a password-protected account with the consumer, they can employ that existing account's authentication mechanisms to verify the consumer's identity. Otherwise, the business is required to verify the requester's identity to a ``reasonable degree of certainty'' by matching either two (before disclosing categories of personal information) or three (before disclosing specific pieces of personal information) data points provided by the consumer with data points maintained by the business.

The CCPA defines a consumer as a California resident ``however identified, including by any unique identifier,''\footnote{Cal. Civil Code \S1798.140(g).} which means that consumers need not use their real names to identify themselves when making VCRs. That is, the CCPA allows consumers to use pseudonyms when transacting with businesses and exercising their privacy rights, and does not require that they divulge their legal names to make VCRs (i.e., for verification, it only needs to match the personal information previously collected by the business).

\subsection{Comparison with the GDPR}
The EU's General Data Protection Regulation (GDPR), which went into effect on May 25, 2018, is considered to be one of the most comprehensive data protection laws to date~\cite{Burgess2020}. Similar to the CCPA, the GDPR offers strong privacy protections to individuals and imposes obligations on businesses conducting business in Europe. In particular, the GDPR also requires companies to disclose their data collection and sharing practices in a privacy policy and respect individuals' right to be informed and right of access to personal information pertaining to them. 

Despite the similarities in the rationale between the CCPA and GDPR, there are also important differences with regard to the scope and application of specific provisions~\cite{Jehl2018, DataGuidanceGDPR}:

\begin{enumerate}
  \item \textbf{Personal Scope.} The GDPR applies broadly to entities that establish the means and purposes of the processing of Europeans' personal information, covering natural and legal persons, for-profit, non-profit, and public entities, small and large organizations, irrespective of their size or revenue. On the other hand, the CCPA only applies to for-profit businesses subject to the criteria enumerated in Section 2.1.
  \item \textbf{Material Scope.} The CCPA excludes specific categories of personal information from its scope of application covered by industry-specific federal privacy laws, whereas the GDPR does not feature such exceptions. For instance, medical information covered by the Health Insurance Portability and Accountability Act (HIPAA) and financial information covered by the Gramm-Leach-Bliley (GLB) Act are both outside of the scope of application of the CCPA.
  \item \textbf{Required Notices.} The CCPA requires covered businesses to disclose in their privacy policies the categories of personal information collected, sold, or disclosed for a business purpose in the preceding 12 months.
  \item \textbf{Right of Access.} The CCPA mandates that companies provide personal information requested by the consumer under the right to know ``in a portable and, to the extent technically feasible, readily usable format that allows the consumer to transmit this information to another entity without hindrance,'' effectively establishing the right to data portability. In contrast, the GDPR separates the right of access and the right to data portability, which have their own conditions. 
  \item \textbf{Procedures.} The CCPA requires organizations to respond to consumers' request in 45 days starting with the receipt of the request, extendable once by an additional 45 days. The GDPR requires covered entities to respond within one month, extendable once by an additional two months. 
  \item \textbf{Penalties.} The GDPR empowers competent data protection authorities to both assess any violations of the law and directly issue fines to entities. In contrast, the Attorney General of the State of California is responsible for assessing violations of the CCPA and bringing civil actions against the offending businesses to seek statutory damages in court.
\end{enumerate}

The next section provides an overview of prior studies investigating the efficacy of the right of access, primarily under the GDPR. We believe that although the methodologies and general findings are applicable to our study, the highlighted differences between the two data protection laws also necessitate the present exploration of businesses' compliance with the CCPA. 

\subsection{Efficacy of Subject Access Requests}

Our work relates to prior studies that investigated how effective subject access request (SAR) mechanisms are in helping data subjects exercise their rights~\cite{veys2021pursuing, di2019personal, di2022revisiting, kroger2020app, adhatarao2021ip, alizadeh2020gdpr, bufalieri2020gdpr, urban2018unwanted}. In~\cite{urban2019study}, SARs were sent to 38 third-party businesses in an effort to evaluate how they comply with Article 15 of the GDPR, and the study showed that most failed to properly disclose all relevant user data in their responses to the requests. Urban et al.~\cite{urban2018unwanted} sent SARs to 36 organizations and found that 58\% delayed responding to the requests. Kr\"oger et al.~\cite{kroger2020app} sent similar requests to app developers over a period of a few years and identified potential weaknesses in the processes developers followed to handle and respond to such requests, which continued to exist even after GDPR became enforceable. Similarly, the results of sending SARs to businesses in~\cite{ausloos2018shattering} highlighted the difficulty they experienced finding all data needed to respond to the requests. The authors also emphasized the importance of using automation whenever possible when responding to SARs and developing templates that businesses can follow so that they can reach a state of ``legal certainty,'' where they can be assured that they are in compliance with laws that provide users with the right to access their data. Tolsdorf et al.~\cite{tolsdorf2021case} identified data incompleteness and inconsistency issues when evaluating the accuracy of information displayed in privacy dashboards for a number of service providers.   

Herrmann and Lindemann~\cite{herrmann2016obtaining} observed that businesses were more likely to respond to data deletion requests than subject access requests, and identified websites that adopted SAR mechanisms that made them vulnerable to revealing their users' data in their responses to adversarial data access requests. In a number of other studies, researchers further examined how businesses' SAR mechanisms can be used by adversaries to extract subjects' personal data through social engineering attacks (e.g., impersonation)~\cite{di2019personal, di2022revisiting, pavur2019gdparrrrr, cormack2016subject, boniface2019security, cagnazzo2019gdpirated}. Di Martino et al.~\cite{di2019personal} showed how these types of attacks can be mounted against a number of organizations by relying on information that is available to the public. In their follow-up work~\cite{di2022revisiting}, they proposed alternative approaches to authenticating data subjects that can help businesses strengthen their SAR mechanisms by reducing the likelihood of leaking subjects' personal data when responding to data access requests made by adversaries. Jordan et al.~\cite{jordan2021viceroy} focused specifically on addressing the problem of how organizations can respond to data access requests that do not have corresponding user accounts.

While prior work has investigated organizations' responses to SARs from a number of different perspectives, we believe that the literature is yet to paint a complete picture on the extent to which responses to SAR are consistent with disclosures made in privacy policies and actual system behaviors. Researchers investigated whether SAR processes are sufficiently explained in privacy policies or aligned with the requirements of applicable laws and compared privacy policy disclosures to responses to SARs~\cite{urban2019study,boniface2019security,bufalieri2020gdpr,bowyer2022human,urban2018unwanted}, but we are unaware of studies that compared organizations' responses to actual system behaviors. We systematically compare information obtained from the three sources of information we considered: privacy policies, responses to SARs and actual app behaviors.   

Researchers also studied the usability of subject access request and deletion mechanisms from a number of different angles, including the ease of initiating the requests as well as the extent to which the content of the responses can be understood by average users~\cite{veys2021pursuing, wei2020twitter, habib2020s, bowyer2022human, urban2019your}. After investigating users' awareness of their rights under the GDPR in~\cite{kuebler2021right}, researchers found that users do not have sufficient understanding of their ``right to data portability.''
Habib et al.~\cite{habib2020s} uncovered challenges users experience with locating information related to how to exercise their privacy rights and correctly using the privacy controls made available to them by businesses. Veys et al.~\cite{veys2021pursuing} observed how real users interacted with the content of the responses obtained from businesses after requesting to download their data. They found that most responses are yet to be considered \textit{accessible} to users and identified areas where future improvements can be made to better align these responses with user expectations~\cite{veys2021pursuing}. Urban et al.~\cite{urban2019your} highlighted the importance of improving the designs of current user-facing tools provided by organizations to allow users to understand how their data is used.
After studying the extent to which responses to SARs submitted to Twitter are empowering real users to understand how their data was used in ad targeting, Wei et al.~\cite{wei2020twitter} similarly found content-related issues that might negatively affect how understandable and readable ad explanations are to users. Table~\ref{tab:comparison} compares some of the key metrics of this study with those of prior work.

\begin{table}
\small\centering
\caption{Comparison of Key Metrics with Related Work}
\label{tab:comparison}
\begin{tabular}{l|l|l|l|l|l}
\textbf{Study}               & \begin{tabular}[c]{@{}l@{}}\textbf{Request }\\\textbf{Count}\end{tabular} & \begin{tabular}[c]{@{}l@{}}\textbf{Response }\\\textbf{Count}\end{tabular} & \begin{tabular}[c]{@{}l@{}}\textbf{GDPR or}\\\textbf{CCPA?}\end{tabular} & \begin{tabular}[c]{@{}l@{}}\textbf{Policy }\\\textbf{Analysis?}\end{tabular} & \begin{tabular}[c]{@{}l@{}}\textbf{App }\\\textbf{Analysis?}\end{tabular}  \\ 
\hline
\textbf{This}                & \textbf{109}                                                              & \textbf{80 (73\%)}                                                         & \textbf{CCPA}                                                            & \textbf{Yes}                                                                 & \textbf{Yes}                                                               \\
\cite{adhatarao2021ip}       & 109                                                                       & 62 (57\%)                                                                  & GDPR                                                                     & No                                                                           & No                                                                         \\
\cite{urban2018unwanted}     & 36                                                                        & 32 (89\%)                                                                  & GDPR                                                                     & Yes                                                                          & No                                                                         \\
\cite{cagnazzo2019gdpirated} & 14                                                                        & 14 (100\%)                                                                 & GDPR                                                                     & No                                                                           & No                                                                         \\
\cite{bufalieri2020gdpr}     & 326                                                                       & 212 (65\%)                                                                 & GDPR                                                                     & Yes                                                                          & No                                                                         \\
\cite{pavur2019gdparrrrr}    & 150                                                                       & 112 (75\%)                                                                 & GDPR                                                                     & No                                                                           & No                                                                         \\
\cite{kroger2020app}         & 225                                                                       & \begin{tabular}[c]{@{}l@{}}43--58 \\(19--26\%)\end{tabular}                & GDPR                                                                     & No                                                                           & No                                                                         \\
\cite{ausloos2018shattering} & 60                                                                        & 44 (73\%)                                                                  & ---                                                                      & Yes                                                                          & No                                                                         \\
\cite{di2022revisiting}      & 40                                                                        & 34 (85\%)                                                                  & GDPR                                                                     & No                                                                           & No                                                                         \\
\cite{di2019personal}        & 55                                                                        & 51 (93\%)                                                                  & GDPR                                                                     & No                                                                           & No                                                                         \\
\cite{urban2019study}        & 38                                                                        & 16 (42\%)                                                                  & GDPR                                                                     & Yes                                                                          & No                                                                         \\
\cite{herrmann2016obtaining} & \begin{tabular}[c]{@{}l@{}}150 apps\\120 sites\end{tabular}               & 43\%                                                                       & GDPR                                                                     & No                                                                           & No                                                                        
\end{tabular}
\end{table}

\subsection{Analysis of Privacy Policy Disclosures}
Others have focused on understanding apps' and websites' privacy practices by analyzing disclosures made in privacy policies~\cite{harkous2018polisis, andow2020actions, wang2018guileak, zimmeck2019maps, zimmeck2017automated}. Some proposed systems, such as POLICHECK~\cite{andow2020actions}, MAPS~\cite{zimmeck2019maps} and HPDROID~\cite{fan2020empirical}, which automated the process of comparing disclosures made in privacy policies about how user data is used, collected, or shared with personal data transmissions observed as a result of performing technical analyses~\cite{wang2018guileak, andow2020actions,zimmeck2019maps, zimmeck2017automated, slavin2016toward}. The literature also proposed systems, such as Polisis~\cite{harkous2018polisis}, PI-Extract~\cite{bui2021automated} and PrivacyFlash~\cite{zimmeck2021privacyflash}, which made it possible to transform privacy policies into formats that are more understandable to users or auto-generate policies that reflect actual app behaviors. Linden et al.~\cite{linden2018privacy} found that disclosures made in privacy policies improved as a result of GDPR enforcement, but that more improvements would have to be made before they can be considered usable and transparent to users. Other recent studies have also examined the accuracy of disclosures made in privacy policies~\cite{andow2019policylint, okoyomon2019ridiculousness, wang2018guileak}. Compared to prior studies, we follow a systematic approach to analyzing apps' privacy policies by having coders answer a set of questions that are reflective of the requirements of the CCPA.


\subsection{Data Practices of Mobile Apps}
Finally, our work also relates to prior studies that investigated potential privacy violations in online systems, including mobile apps and websites~\cite{nguyen2021share, reyes2018won, han2020price}. To examine the extent to which apps comply with privacy regulations, researchers relied on static and dynamic app analysis tools to identify potential legal violations at scale~\cite{reyes2018won, han2020price, feal2020angel, nguyen2021share, ren2018longitudinal, jia2019leaks}. These studies identified a range of deceptive data collection and transmission practices and highlighted the need for stronger enforcement actions by regulators. While prior work has examined apps' level of compliance with privacy regulations by looking into network flows, privacy policy disclosures, or responses to data subject access requests, our investigation compares the data obtained across all three of these sources to evaluate the effect of the CCPA on developers' privacy practices. We believe that our work, therefore, is crucial in evaluating the overall efficacy of CCPA and its utility to mobile app users.

\section{Methodology}
\label{sec:methods}

We aim to uncover contradictions between personal information...

\begin{enumerate}
    \item that we record being collected and transmitted by an app using dynamic and static analysis;
    \item disclosed to us in response to a ``right to know'' request we made after using the mobile app; and
    \item that the app developer claimed to collect in their app's privacy policy.
\end{enumerate}

The following sections cover each part of the study in more detail. We additionally describe our procedure for selecting the Android apps that we examined, as well as our procedure for testing the apps and submitting the verifiable consumer requests.

\subsection{Dataset}
We focused on the 8 top-ranked Android mobile apps in the 20 Google Play Store categories that have the highest number of cumulative app installs. Companies developing these apps fall or can be reasonably inferred to fall under the CCPA definition of a ``business.''\footnote{A ``business'' includes mobile app developers that are for-profit entities and conduct business in California (i.e., make their applications available in California) and meet at least one of the following three criteria: (1) collect the personal information of at least 50,000 consumers in California; (2) have an annual gross revenue in excess of \$25 million, or (3) derives 50 percent or more of its annual revenues from selling California consumers’ personal information (CCPA, 1798.140(c)).} We selected only one mobile app (with the highest user install count) per developer in order to have the ability to match the personal information disclosed by the developer with the app that we tested and to examine a broader range of developer practices for responding to VCRs.

Furthermore, we replaced certain apps that we were unable to test. This included apps, for instance, that required business accounts, financial information, or additional hardware devices. This selection procedure produced a total of \totalAppCount{} unique apps, which we downloaded with their privacy policies in November 2021.

It is important to note that, although our procedure was designed to select developers that we expected to be covered by the CCPA, the resulting list was only an approximation (i.e., we could not be sure that all of these developers were \textit{actually} subject to the CCPA), as we used the number of app installs to gauge the total number of California consumers from whom an app may have collected personal information. Nonetheless, we could not be sure, and as an ethical matter, we did not want to waste people's time by submitting VCRs to organizations that were not required to respond to them. Thus, we further limited our study to only those companies that explicitly mentioned CCPA in their privacy policies. Under the FTC Act\footnote{15 U.S.C. \S45.} (and various other state consumer protection laws), businesses in the U.S. are prohibited from materially misrepresenting their practices to consumers. This includes making false statements in privacy policies, which the FTC enforces (e.g., \cite{FTCFlo2021}). Thus, any business that states in their privacy policy that they respond to CCPA VCRs must actually do so, regardless of whether or not they are \textit{actually} covered by the CCPA.

Two researchers from our team independently read the text of \totalAppCount{} privacy policies to determine whether or not each contained references to the CCPA. For cases without a majority consensus, a third researcher provided the tie-breaking vote. Our analysis indicated that out of the selected \totalAppCount{} apps, \totalCCPAAppCount{} (\totalCCPAAppPercentage{}) include CCPA-specific disclosures in their privacy policies (with Krippendorff's alpha = \krippendorffCCPA{}, indicating an acceptable level of inter-rater agreement~\cite{krippendorff2011computing}). For the remainder of this paper, our discussion will focus primarily on these \totalCCPAAppCount{} apps.

\subsection{App Analysis}

We used an instrumented version of Android 9.0 (Pie) that monitored resource accesses (e.g., access to Android APIs) and logged all network traffic, regardless of the use of TLS. (Prior published work has applied a similar approach~\cite{reardon201950,Sanfilippo2020,andow2020actions,han2020price,alomar2022developers}.) Because network traffic was captured at the OS level (as opposed to using a proxy), we were still able to observe and decrypt transmissions that were secured using certificate pinning. Since the values of identifiers (and other personal information) were known for each device, our tools automatically searched for various permutations in the captured network traffic, including hashes (e.g., MD5, SHA-1, SHA-256, etc.).

Using this instrumentation on Google Pixel 3a devices, we automatically recorded decrypted network traffic, which included destinations (i.e., hostname, port, IP) and payloads. Decrypted traffic payloads included API endpoints and key/value pairs. All network traffic was attributed to specific apps and their SDKs, using a combination of kernel-level instrumentation to attribute sockets to processes and stack inspection to identify specific SDKs. A variety of open source tools for collecting network traffic can be used to verify our results and, we believe, reproduce our findings from scratch (e.g., ~\cite{Frida, Zap}). While the instrumentation was specifically written for Android Pie (9), which was released roughly three years prior to our testing, millions of people still use Pie (e.g., at the time that we conducted our study, roughly 20\% of US Android users were using Pie or earlier~\cite{AndroidMarket2021}), many with CCPA rights. We also have no reason to believe that the same app binaries would be more/less compliant under newer Android versions.


\paragraph{\textbf{Pseudonyms.}}
Similar to~\cite{zang2015knows}, we generated pseudonyms and other fictitious values for different types of personal information covered by the CCPA to facilitate the subsequent search for this data in the logs produced by app testing and to improve the ecological validity of our study. Our motivation behind using ``fake'' data was to reduce the number of confounding variables: while all experimenters were California residents, if we used our real names and identifiers, we would not know whether data received from CCPA VCRs was collected by the company during the study period or before (or possibly from other sources).


The CCPA defines a consumer as a California resident ``however identified, including by any unique identifier,''\footnote{Cal. Civ. Code \S1798.140(g).}, therefore, the usage of fictitious data did not legally affect the requirement of the companies to respond to our requests. This provision ensures that companies that only collect pseudonyms are still subject to CCPA requests, while also disincentivizing companies from collecting additional personal information solely for the purpose of responding to requests. A physical address (and email, phone number, etc.) can be fictitious, so long as they can be used to identify the California consumer who is the data subject. Thus, the use of pseudonyms both reduced confounding factors and was legally valid.


We produced pseudonymous data using random value generators, such as the Random Lists~\cite{randomlists} website and Faker Python package~\cite{faker}. We obtained other types of personal information, including device identifiers and geolocation data, directly from our test devices. We present our data taxonomy in Appendix~\ref{sec:data_taxonomy}, while Table~\ref{tab:taxonomy} provides examples of personal information that we used.
  
\paragraph{\textbf{Testing Procedure.}}
We manually tested the selected 109 apps, each for approximately 15-20 minutes using test phones with our instrumented version of the Android operating system. We set up each test phone---to be used by an individual tester in California---to use its own set of pseudonymous identifiers, such as the phone number, email address, usernames, and other types of information. During each test, we created a user account for the app (if applicable) and input the predefined pseudonymous data corresponding to the specific test phone, as described above. We later searched for the predefined data values within the resulting test logs (which included captured network traffic), as well as performed an open-ended search to see if the app transmitted other personal data. 

\paragraph{\textbf{Data Recipients.}}
Apps can transmit data both to first- and third-party destinations in order to deliver essential and non-essential functionality. Specifically, we might observe an app transmit the same personal information only to domains controlled by the app developer or to a combination of first- and third-party endpoints.

First, we categorized the observed destination domains as either first- or third-party for each tested app. Using the same approach as in~\cite{trimananda2022ovrseen}, we tokenized the destination domain and the app package name. We then classified a specific domain as first party if its tokens appeared in the app's privacy policy URL or matched the package name's tokens, otherwise, we labeled the domain as third party. Next, we went over the resulting party labels for each domain and manually corrected any mistakes. For each third-party domain, we also obtained the effective second-level domains (eSLD) using \texttt{tldextract} and used it to locate the entity that controls it using Crunchbase, Netify, and other online resources. Two researchers from our team assigned a category to each third-party domain using the information that we obtained from our online search, which we then used to compare against the categories of recipients in VCR responses and privacy policy disclosures. 

The CCPA recognizes that a first party can either directly or indirectly collect personal information.\footnote{Cal. Civ. Code \S1798.130(a)(3)(A).} As such, the collection of personal information via third parties (either service providers or third parties under the CCPA) still triggers the CCPA obligations on the first party as if the app developer directly collected the personal information itself. The liability of first parties for third-party app and website data collection has been affirmed by People of the State of California v. Sephora USA, Inc.~\cite{californiaSephora}.

For this reason, we labeled each data point (e.g., for purposes of Table~\ref{tab:undisclosedPII}) that we observed being captured and transmitted to a third-party domain (e.g., using SDKs, codebases, or other pieces of code in the app) as collected both by the first party (i.e., the app developer) and the third party. We categorized the data point as collected by the first party if the app transmitted it only to domain(s) controlled by the app developer.   


\subsection{Verifiable Consumer Requests}
For each tested app, we identified directions in its privacy policy for how to submit a verifiable consumer request (VCR). To avoid abusing the time and resources of developers who do not have to comply with the provisions of the CCPA, we erred on the side of caution and only submitted verifiable consumer requests to developers who explicitly referenced the CCPA in their privacy documents.

As part of each request, we asked to obtain all types of information that a business is required to provide under the CCPA in response to a consumer request: 

\begin{enumerate}
    \item specific pieces and categories of personal information requested, collected, and shared by the app;
    \item categories of sources from which the personal information was collected;
    \item business or commercial purposes for collecting the personal information; and 
    \item specific names and categories of third parties with whom the app developer shared personal information.\footnote{CCPA does not require businesses to disclose specific third parties, however, some app developers opt in their privacy notices to provide that information upon request.}
\end{enumerate}

We submitted each request from the same pseudonymous email account that was used to test the app. We employed email templates to ensure a level of uniformity when, for instance, we submitted the initial requests, sending follow-ups if the developer did not respond, asking for an alternative identity authentication mechanism, etc. We provide the email templates that we used to submit the requests and follow up with the developer in Appendix~\ref{sec:vcr_emails}. Nevertheless, some developers still instructed us to use an alternative method for submitting the request, such as a privacy management platform.




\subsection{Privacy Policy Analysis}
Additionally, we analyzed disclosures made in the privacy policies of tested apps using a deductive approach to qualitative coding. Our codebook contains codes for the collection and sharing of categories of personal information taken from Cal. Civil Code 1798.140. One of the authors with experience assisting companies in complying with the CCPA requirements developed the codes for the categories of third parties. We include the resulting codebook, code descriptions, and prompts in Appendix~\ref{sec:codebook}.

As discussed previously, we first identified whether each policy contained references to the CCPA using the following prompt: ``\textit{Does this app developer include disclosures that reference the CCPA, either as part of the general privacy policy or as a standalone document?}'' We then analyzed each of the \totalCCPAAppCount{} privacy policies containing CCPA-specific information to identify information about the developer's data collection and sharing practices. In particular, for each category of personal information defined under the CCPA (e.g., identifiers or geolocation), we examined whether an app developer collected or disclosed each category and to which category of recipients. 

At least two annotators from our team first independently located the relevant privacy policies, and then used the prompts enumerated in Table~\ref{tab:policyAnalysis} to locate the disclosures that pertained to the collected and shared categories of personal information and the categories of third parties. We then computed Krippendorff's $\alpha$ to evaluate the inter-rater reliability on a per-question basis~\cite{krippendorff2018content}. We resolved any divergences in our responses using a majority vote or, if a majority was absent, a third researcher independently provided the tie-breaking vote. After resolving the disagreements, we obtained a list of categories of personal information and recipients that we compared against our app analysis results. 

\subsection{Comparison}
We compared these three data viewpoints to quantify the accuracy and completeness of the information disclosed by the developers.
We first compared each specific piece of personal information that we observed being collected and shared with the specific pieces of information disclosed by the developer in the VCR, when applicable. In this case, we simply matched the values that we observed being collected and shared with the values provided to us by the developer. As mandated by the CCPA, we only accepted responses containing the values (and not just the types of information) to be valid with respect to disclosing the specific pieces of personal information.

Furthermore, we compared the categories of collected and shared information and the categories of recipients that we observed during app testing with the the same categories disclosed in the VCR and privacy policy. We only considered the categories disclosed in the VCR responses to be valid if we were able to sufficiently match them with the CCPA-defined categories of personal information and to our categories of third parties. These categories included common types of recipients that we observed across different app privacy policies and VCR responses, such as advertising networks, marketing partners, analytics providers, fraud and security, search engines, social media networks, payment processors, customer support providers, storage and infrastructure, affiliates, and law enforcement. We obtained the same categories of personal information and third parties from the privacy policies using the qualitative coding approach discussed previously. The CCPA-defined categories of personal information as well as the categorization of our own PII types are presented in Table~\ref{tab:taxonomy}.

Once we had obtained these categories, we identified the categories that the developer had collected but not disclosed by looking at the difference between categories that we observed during app testing and the categories provided by the app developer in the privacy policy and VCR.

\subsection{Ethics}
We performed a study of institutional processes and did not collect data \textit{about} individuals~\cite{OHRP}. As such, our IRB determined that our study did not meet the legal definition of human subjects research, and therefore declined to review it. We nonetheless spent over a year deliberating how to conduct it ethically, including avoiding guessing whether a company was subject to CCPA, not incurring costs by asking legal questions, and making sure correspondence was not perceived as legal threats, ethical issues that have come up for other researchers~\cite{Mayer2021}. Instead, we performed a measurement study of publicly-available services by exercising our legal rights using the methods companies themselves prescribed.

We acknowledge that some companies may not have automated systems to process CCPA requests, and therefore processing our VCRs may have imposed costs on them. However, we believe that business' interests in this regard are outweighed by the public interest in understanding CCPA effectiveness. This is also a straw man argument: all individuals who made CCPA requests for our study were legitimately interested in learning about companies' privacy practices and made legally-valid requests to do so; that they additionally followed a prescribed methodology and shared the results for research purposes does not suddenly make the requests invalid or unethical. CCPA empowers California residents with \textit{rights}, which must be honored regardless of intent.
\section{Analysis}
\label{sec:analysis}
We present the results from submitting the VCRs, focusing on the methods available to do so, the types of information required to initiate and verify requests, and the percentage of developers who completed the requests, with an emphasis on the disclosure of the CCPA-specific information, as enumerated in Section 3.4. Furthermore, we compare the personal information provided to us by the developers with our dynamic analysis of their Android apps.

\subsection{Access Requests}
We analyzed the \totalCCPAAppCount{} apps with CCPA-specific information in their privacy policies. Whenever possible, we created an account with each app using an email address created specifically for this study and unique to the testing phone. As a result, we registered accounts with \accountHolderAppCount{} (\accountHolderAppPercentage{}) apps. 

The majority of developers (\twoOrMoreSARMethodsAppPercentage{}) provided at least two methods for submitting the VCR. The most common method was by email, with \emailMethodCount{} (\emailMethodPercentage{}) companies offering it as an option. The next most common method was a dedicated VCR form or portal offered by \totalSARPortalMethodCount{} (\totalSARPortalMethodPercentage{}) companies. Notably, \oneTrustMethodCount{} of these companies relied on OneTrust~\cite{one-trust}, a third-party suite of products that includes support for SAR management, with the remaining \ownSARPortalMethodCount{} either relying on another third-party provider or implementing their own solutions. We identified a number of other methods for submitting requests, including a phone number (\phoneMethodPercentage{}), contact via customer support service (\customerSupportMethodPercentage{}), physical mail (\mailMethodPercentage{}), account or in-app privacy settings (\settingsMethodPercentage{}) or through a Google Form (\googleFormMethodPercentage{}).

\begin{table}[t]
\small
\caption{Distribution of Methods for Submitting VCRs}
\label{tbl:methods}
\begin{tabular}{lll}
\hline
Name of Method                       & Count & Proportion \\ 
\hline\hline
Email & \emailMethodCount{}    & \emailMethodProportion{}     \\
Company DSAR Portal & \ownSARPortalMethodCount{}    & \ownSARPortalMethodProportion{}     \\
Phone                                & \phoneMethodCount{}    & \phoneMethodProportion{}     \\
Customer Support Service             & \customerSupportMethodCount{}    & \customerSupportMethodProportion{}      \\
Physical Mail                        & \mailMethodCount{}    & \mailMethodProportion{}      \\
OneTrust DSAR Portal                 & \oneTrustMethodCount{}    & \oneTrustMethodProportion{}     \\
Account Privacy Settings             & \accountPrivacyMethodCount{}    & \accountPrivacyMethodProportion{}     \\
In-App Privacy Settings              & \appPrivacyMethodCount{}     & \appPrivacyMethodProportion{}     \\
In-App Feedback Form     & \appFeedbackMethodCount{}     & \appFeedbackMethodProportion{}     \\
Google Form                          & \googleFormMethodCount{}     & \googleFormMethodProportion{}      \\ \hline
\end{tabular}
\end{table}

Whenever possible, we submitted VCRs using email or a customer support service. In these cases, our messages to the companies included a self-attestation of California residence, as well as the pseudonyms and email addresses associated with the phone used for testing the app. However, in \alternativeRequestCounts{} cases, the app developer directed us to use an alternative VCR submission method other than the one we had chosen. Ultimately, we submitted \emailRequestCounts{} VCRs via email and \CustomerSupportRequestCounts{} VCRs via customer support or a feedback form out of the total \totalCCPAAppCount{} requests sent out. We submitted the remaining \PreparedFormRequestCounts{} requests either using a provided VCR portal (\DSARRequestPercentage{}) or within an app's or account's privacy settings (\SettingsRequestPercentage{}).

When using a dedicated form, portal, or in-app privacy controls to submit the VCR, we generally received a confirmation of the request within the same user interface. For this reason, we focused on the \FreeFormRequestCounts{} apps that required a free-form request submission to see if the companies would confirm the receipt of our request within the statutory 10 day period mandated by the CCPA. Out of these \FreeFormRequestCounts{} companies, \FreeFormConfirmedReceiptCounts{} (\FreeFormConfirmedReceiptPercentage{}) explicitly confirmed our request, whereas the remaining \FreeFormNotConfirmedReceiptCounts{} (\FreeFormNotConfirmedReceiptPercentage{}) did not. 

Companies also must verify the identity of consumers submitting VCRs to ensure they do not inadvertently disclose personal information to someone impersonating the data subject. Therefore, we also recorded information that we provided or any authentication steps we performed to verify our VCR (Table \ref{tab:analysis}). We implicitly verified the ownership of our email address in \emailRequestCounts{} instances, when we made the request via email. For all other cases, \NotEmailMethodVerifiedEmailYesCount{} companies requested email verification after submitting the request, typically by clicking a link or providing a unique PIN sent to the testing email address. Furthermore, \VerifiedAccountYesCount{} companies required us to successfully log into our accounts either to submit or to verify the VCR. 

App developers also requested specific pieces of personal information to match against their records, either as part of the initial VCR submission process or by following up with us after we submitted our requests. Most often, developers asked us to provide some basic information about ourselves, including, our email address (\VerificationEmailCount{} instances), full name (\VerificationNameCount{}), state (\VerificationStateCount{}), and country of residence (\VerificationCountryCount{}). Developers also requested technical information that is not always easily accessible for smartphone users. In particular, we were asked to provide the Android Advertising ID (AAID) in \VerificationAAIDCount{} cases, a company-defined `device` or `user` ID in \VerificationDeviceUserIDCount{} cases, and our current IP address in \VerificationIPCount{} cases. Table \ref{tbl:methods} presents a breakdown of the different types of information or actions required to verify the VCRs.  

Some companies had more stringent requirements to complete their identity verification, either at the moment or after submitting the VCR. Five companies out of \totalCCPAAppCount{} required us to certify the accuracy of the provided information under penalty of perjury and \VerificationAffidavitCount{} required a signed affidavit that, in at least one case, had to be notarized.\footnote{This is explicitly prohibited by regulations (\S999.323(d)).} Furthermore, two companies requested proof of phone number ownership by providing a recent mobile operator bill, another two asked for photocopies of a government-issued ID, and one company outsourced identity verification to the ID.me service, which allows an individual to verify themselves either by providing a photocopy of their government-issued ID or their phone number to allow a look-up with the mobile operator records. Finally, one company asked us to ``make [ourselves] available for a phone call with a [redacted] customer service representative who will call from [their] privacy line.'' In these instances when we could not furnish such documents, we requested an alternative verification method through logging into our account and providing details of that login to the company, if applicable. The CCPA regulations explicitly provide for such an alternative verification method for account-holders. Two companies agreed, and allowed us to verify our identity using the alternative verification method.

The majority of companies, namely \CAResidencyNoCount~or \CAResidencyNoPercentage, did not ask for proof of our California residency. Out of the remaining \CAResidencyYesCount~app developers, three asked us to provide proof of our address (e.g., a bank statement or a recent utility bill), one requested a government-issued ID showing California residency (e.g., a California driver's licence), one asked us to sign a declaration of California residency under the penalty of perjury, and the remaining two requested California state residency verification via ID.me and the phone call, as described previously. 

\begin{table}[t]
\small\centering
\caption{Methods or Information Required to Verify VCR}
\label{tab:analysis}
\begin{tabular}{ll}
\hline
Method or PII Type & Count \\ 
\hline
\hline
Email                                         & \VerificationEmailCount{}                                                   \\
Account Authentication                        & \VerifiedAccountYesCount{}                                                   \\
Email Authentication                          & \NotEmailMethodVerifiedEmailYesCount{}                                                    \\
Full Name                                     & \VerificationNameCount{}                                                   \\
State of Residence                            & \VerificationStateCount{}                                                   \\
App-specific Information                      & \VerificationAppSpecificInfoCount{}                                                    \\
Country of Residence                          & \VerificationCountryCount{}                                                   \\
Username                                      & \VerificationUsernameCount{}                                                     \\
Phone Number                                  & \VerificationPhoneNumberCount{}                                                           \\
Postal Address                                & \VerificationPostalAddressCount{}  \\
Device or User ID                             & \VerificationDeviceUserIDCount{}                                                    \\
Android Advertising ID (AAID)                 & \VerificationAAIDCount{}                                            \\
Certification w/ Penalty of Perjury    & \VerificationCertificationPerjuryCount{}                                     \\
Signed Affidavit                              & \VerificationAffidavitCount{}                                                    \\                                                 
Photocopy of a Government-issued ID           & \VerificationGovernmentIDCount{}   \\
Phone Authentication                          & \VerificationPhoneNumberOwnershipCount{}    \\
Current IP Address                            & \VerificationIPCount{}  \\
Date of Birth                                 & \VerificationDOBCount{}     \\
ID.me                                         & \VerificationIDMeCount{}    \\
Call with a Company Representative            & \VerificationCallCount{}    \\
\hline
\end{tabular}
\end{table}

\subsection{Developer Responses}
Out of the \totalCCPAAppCount{} requests that we sent out, we did not receive a response from the developer in \NoResponseCount{} (\NoResponsePercentage{}) cases. In these instances, the developer either did not respond to the initial request or became unresponsive after a brief interaction, for instance, after asking for verification. In all of these cases, we followed up with the app developers at least once to confirm that they were unresponsive. 

We were unable to verify our identity to the company's satisfaction in \NoResponseCannotVerifyCount{} (\NoResponseCannotVerifyPercentage{}) other cases, as we were unable to produce the requested documentation and the company did not agree to use an alternative method. Finally, \ResponseCannotComplyCount{} (\ResponseCannotComplyPercentage{}) developers could not verify our identity to a sufficient degree and, thus, did not respond with any personal information. We excluded these \ResponseNoCount{} cases from our analysis of the responses and focused on the remaining \ResponseYesCount{} responses.


\paragraph{\textbf{Human vs. Automated Responses}}
We first identified the proportion of companies employing automation when responding to our VCRs. Similar to~\cite{urban2019study}, we labeled responses that directly answered to our questions as ``human.'' In contrast, we marked responses sent by a computer system (e.g., help desk ticketing software) or containing only generic privacy-related information as ``automated.'' Out of \ResponseYesCount{} responses, we labeled \HumanResponseCount{}~(\HumanResponsePercentage{}) responses as ``human'' and the remaining \AutomatedResponseCount{}~(\AutomatedResponsePercentage{}) as ``automated.''

\paragraph{\textbf{Follow-up Actions}}
We first examined the number of actions that the data subject would have to perform to successfully receive a response to their VCR, and the amount of time they would have to wait for the company to reply back. Across the \ResponseYesCount{} responses, we performed an average of \ActionsMean{} (±\ActionsSD{}, median = \ActionsMedian) actions to obtain our VCR response, including submitting the request, passing identity verification, following up with the developer, etc. The most actions that we performed was \ActionsMax{}. Additionally, it took us \DurationDaysMean{} (±\DurationDaysSD{}, median = \DurationDaysMedian{}) days on average to receive responses to our VCRs, however, the average was skewed heavily by developers who instantly replied back with the response (e.g., if made through in-app account settings) and those that took extraordinarily long, with the longest duration to complete the request of \DurationDaysMax{} days.

\paragraph{\textbf{Composition of the Response}}
Out of these \ResponseYesCount{} companies, \ResponseWithDataCount{} (\ResponseWithDataPercentage{}) provided data in response to our request, \ResponseWithoutDataCount{} (\ResponseWithoutDataPercentage{}) replied that they held no data on us and the remaining \ResponseSettingsCount{} (\ResponseSettingsPercentage{}) told us to obtain the requested information directly from our account profile. 

For the \ResponseWithDataCount{} companies that provided us data, we examined whether they provided all types that a business is required to provide under the CCPA (Section 3.4). All but one app developer provided us with specific pieces of information in their responses. However, compliance with other parts of the CCPA's right to know was less uniform. For instance, only \collectedCategoriesPIIYesCount{} (\collectedCategoriesPIIYesPercentage{}) companies provided the categories of personal information collected from us, \disclosedCategoriesPIIYesCount{} (\disclosedCategoriesPIIYesPercentage{}) provided the categories of personal information disclosed or sold to third parties, \partyCategoriesYesCount{} (\partyCategoriesYesPercentage{}) provided the categories of those third parties, \purposeYesCount{} (\purposeYesPercentage{}) responded with the business or commercial purpose for collecting or selling our personal information, and \sourcesYesCount{} (\sourcesYesPercentage{}) disclosed the sources, from which our information was collected. 

\paragraph{\textbf{Compliance}}
The relatively high compliance with the request to provide specific pieces of information is not surprising, as many app developers are likely using tools to automatically respond to CCPA (and GDPR) requests by integrating with and pulling data from their internal customer relationship management (CRM) platforms. Furthermore, in most cases, even when an developer provided the categories of collected or shared personal information or the categories of third parties, sources, or purposes, these disclosures came directly from their privacy policies. We mark these cases as valid disclosures, as we are unable to verify whether those categories in fact apply to our case or not from the developer's response alone.

\paragraph{\textbf{Response Format}}
The \ResponseWithDataCount{} companies that replied with the personal information collected about us communicated this information to us in a number of ways, including \emailResponseCount{} (\emailResponsePercentage{}) companies that included the data directly in the email reply or as an email attachment, \dsarResponseCount{} (\dsarResponsePercentage{}) that provided the data as an attachment on the VCR platform, and \settingsResponseCount{} (\settingsResponsePercentage{}) that made it available from account or in-app privacy settings. The remaining \remainingResponseCount{} companies used a variety of methods to transmit the data to us, including, as a file shared with us via a cloud storage provider, as a download link in the email reply, or via a message sent to us through a customer support portal.

\paragraph{\textbf{Security of the Process}}
We looked at the security mechanisms (if any) used by the developers of the \ResponseWithDataCount{} apps to securely communicate our personal information to us, beyond our email provider's access controls. At least \accessExpiresYesCount{} companies used an expiration time on the download links or files that they shared with us, ranging anywhere from 24 hours to 90 days. However, in 4 of these cases, we verified that the files remained accessible and downloadable even after the stated expiration time. Additionally, \accessAccountYesCount{} app developers relied on their standard account authentication for access control, \accessConfidentialYesCount{} used Gmail's ``confidential mode'' and \accessOtherYesCount{} relied on other access controls, such as those enforced by cloud storage providers. Additionally, \accessEmailYesCount{} companies required email verification to access and download the file, while \accessPasswordYesCount{} secured the data file by setting a password to open it, which they communicated separately to us. 

\paragraph{\textbf{Data Format}}
We looked at the format and characteristics of the \dataFilesCount~data files that contained specific pieces of collected personal information. Developers relayed the files using a number of formats, including CSV (\dataCSVCount~instances), JSON (\dataJSONCount), PDF (\dataPDFCount), Excel (\dataXLSXCount), and TXT (\dataTXTCount). Only \dataTwoOptionCount~companies presented the same data using two different formats, whereas the remaining \dataOneOptionCount~either used a single format or a combination of several comprising a single data record. 

\subsection{Comparison with App Analysis Results}
We strive to not only to understand the process of submitting a VCR under the CCPA, but also the accuracy of the data provided back to us. We first focus on the \specificPIIYesCount~companies who replied with the specific pieces of personal information. In this case, the response to the VCR included specific values that were collected by the developers, therefore, we simply matched the values from the VCR with the data that we observed being transmitted over the network. 

Only \disclosedProvidedSpecificCount~apps that provided us the specific pieces of personal information fully disclosed the extent of their data collection practices. With respect to the enumerated list of categories of personal information defined by the CCPA, we observed the collection, but not the disclosure, of identifiers by \undisclosedProvidedSpecificIdentifiersCount~apps, geolocation data by \undisclosedProvidedSpecificGeolocationCount~apps, sensory data by \undisclosedProvidedSpecificSensoryCount~apps, customer record information by \undisclosedProvidedSpecificRecordsCount~apps and, to a lesser extent, professional information in \undisclosedProvidedSpecificEmploymentCount~cases, characteristics of protected classifications (e.g., gender or age) in \undisclosedProvidedSpecificProtectedCount~cases, and education information in one case.

\begin{table*}
\begin{threeparttable}
\small
\centering
\caption{Counts of Apps that Collected but not Disclosed Various PII}
\label{tab:undisclosedPII}
\begin{tabular}{l|l|l|l|l|l|l} 
\hline
\textbf{Category}         & \textbf{Subcategory} & \textbf{PII Name}  & \textbf{\#Apps}                              & \textbf{TLS \#}                                                                                 & \textbf{1\textsuperscript{st~}Party \#}                                                                       & \textbf{3\textsuperscript{rd} Party \#}                                                                        \\ 
\hline
Identifiers               & User                 & Username           & \undisclosedDeviceusernameCounts{}           & \undisclosedDeviceusernameTLS{} (\undisclosedDeviceusernamePercentageTLS{})                     & \undisclosedDeviceusernameFirstParty{} (\undisclosedDeviceusernamePercentageFirstParty{})                     & \undisclosedDeviceusernameThirdParty{} (\undisclosedDeviceusernamePercentageThirdParty{})                      \\
                          & Network              & IP Address         & \undisclosedIpaddressCounts{}                & \undisclosedIpaddressTLS{} (\undisclosedIpaddressPercentageTLS{})                               & \undisclosedIpaddressFirstParty{} (\undisclosedIpaddressPercentageFirstParty{})                               & \undisclosedIpaddressThirdParty{} (\undisclosedIpaddressPercentageThirdParty{})                                \\
                          &                      & Router MAC         & \undisclosedRoutermacCounts{}                & \undisclosedRoutermacTLS{} (\undisclosedRoutermacPercentageTLS{})                               & \undisclosedRoutermacFirstParty{} (\undisclosedRoutermacPercentageFirstParty{})                               & \undisclosedRoutermacThirdParty{} (\undisclosedRoutermacPercentageThirdParty{})                                \\
                          &                      & Router SSID        & \undisclosedRouterssidCounts{}               & \undisclosedRouterssidTLS{} (\undisclosedRouterssidPercentageTLS{})                             & \undisclosedRouterssidFirstParty{} (\undisclosedRouterssidPercentageFirstParty{})                             & \undisclosedRouterssidThirdParty{} (\undisclosedRouterssidPercentageThirdParty{})                              \\
                          & Device               & AAID               & \undisclosedAaidCounts{}                     & \undisclosedAaidTLS{} (\undisclosedAaidPercentageTLS{})                                         & \undisclosedAaidFirstParty{} (\undisclosedAaidPercentageFirstParty{})                                         & \undisclosedAaidThirdParty{} (\undisclosedAaidPercentageThirdParty{})                                          \\
                          &                      & Hardware ID        & \undisclosedHardwareidCounts{}               & \undisclosedHardwareidTLS{} (\undisclosedHardwareidPercentageTLS{})                             & \undisclosedHardwareidFirstParty{} (\undisclosedHardwareidPercentageFirstParty{})                             & \undisclosedHardwareidThirdParty{} (\undisclosedHardwareidPercentageThirdParty{})                              \\
                          &                      & IMEI               & \undisclosedImeiCounts{}                     & \undisclosedImeiTLS{} (\undisclosedImeiPercentageTLS{})                                         & \undisclosedImeiFirstParty{} (\undisclosedImeiPercentageFirstParty{})                                         & \undisclosedImeiThirdParty{} (\undisclosedImeiPercentageThirdParty{})                                          \\
                          &                      & IMSI               & \undisclosedImsiCounts{}                     & \undisclosedImsiTLS{} (\undisclosedImsiPercentageTLS{})                                         & \undisclosedImsiFirstParty{} (\undisclosedImsiPercentageFirstParty{})                                         & \undisclosedImsiThirdParty{} (\undisclosedImsiPercentageThirdParty{})                                          \\
                          &                      & SIM ID             & \undisclosedSimidCounts{}                    & \undisclosedSimidTLS{} (\undisclosedSimidPercentageTLS{})                                       & \undisclosedSimidFirstParty{} (\undisclosedSimidPercentageFirstParty{})                                       & \undisclosedSimidThirdParty{} (\undisclosedSimidPercentageThirdParty{})                                        \\
                          &                      & Wi-Fi MAC          & \undisclosedWifimacCounts{}                  & \undisclosedWifimacTLS{} (\undisclosedWifimacPercentageTLS{})                                   & \undisclosedWifimacFirstParty{} (\undisclosedWifimacPercentageFirstParty{})                                   & \undisclosedWifimacThirdParty{} (\undisclosedWifimacPercentageThirdParty{})                                    \\
                          &                      & Fingerprint ID     & \undisclosedFingerprintidCounts{}            & \undisclosedFingerprintidTLS{} (\undisclosedFingerprintidPercentageTLS{})                       & \undisclosedFingerprintidFirstParty{} (\undisclosedFingerprintidPercentageFirstParty{})                       & \undisclosedFingerprintidThirdParty{} (\undisclosedFingerprintidPercentageThirdParty{})                        \\
                          & App                  & Identity ID        & \undisclosedIdentityidCounts{}               & \undisclosedIdentityidTLS{} (\undisclosedIdentityidPercentageTLS{})                             & \undisclosedIdentityidFirstParty{} (\undisclosedIdentityidPercentageFirstParty{})                             & \undisclosedIdentityidThirdParty{} (\undisclosedIdentityidPercentageThirdParty{})                              \\
                          &                      & App Fingerprint ID & \undisclosedAppfingerprintidCounts{}         & \undisclosedAppfingerprintidTLS{} (\undisclosedAppfingerprintidPercentageTLS{})                 & \undisclosedAppfingerprintidFirstParty{} (\undisclosedAppfingerprintidPercentageFirstParty{})                 & \undisclosedAppfingerprintidThirdParty{} (\undisclosedAppfingerprintidPercentageThirdParty{})                  \\
                          &                      & Android ID         & \undisclosedAndroididCounts{}                & \undisclosedAndroididTLS{} (\undisclosedAndroididPercentageTLS{})                               & \undisclosedAndroididFirstParty{} (\undisclosedAndroididPercentageFirstParty{})                               & \undisclosedAndroididThirdParty{} (\undisclosedAndroididPercentageThirdParty{})                                \\ 
\hline
Customer Records          & Customer             & Phone Number       & \undisclosedTelephonenumberCounts{}          & \undisclosedTelephonenumberTLS{} (\undisclosedTelephonenumberPercentageTLS{})                   & \undisclosedTelephonenumberFirstParty{} (\undisclosedTelephonenumberPercentageFirstParty{})                   & \undisclosedTelephonenumberThirdParty{} (\undisclosedTelephonenumberPercentageThirdParty{})                    \\
                          & Contacts             & Name               & \undisclosedContactnameCounts{}              & \undisclosedContactnameTLS{} (\undisclosedContactnamePercentageTLS{})                           & \undisclosedContactnameFirstParty{} (\undisclosedContactnamePercentageFirstParty{})                           & \undisclosedContactnameThirdParty{} (\undisclosedContactnamePercentageThirdParty{})                            \\
                          &                      & Phone Number       & \undisclosedContactnumberCounts{}            & \undisclosedContactnumberTLS{} (\undisclosedContactnumberPercentageTLS{})                       & \undisclosedContactnumberFirstParty{} (\undisclosedContactnumberPercentageFirstParty{})                       & \undisclosedContactnumberThirdParty{} (\undisclosedContactnumberPercentageThirdParty{})                        \\
                          & Residence            & Street             & \undisclosedPostalstreetCounts{}             & \undisclosedPostalstreetTLS{} (\undisclosedPostalstreetPercentageTLS{})                         & \undisclosedPostalstreetFirstParty{} (\undisclosedPostalstreetPercentageFirstParty{})                         & \undisclosedPostalstreetThirdParty{} (\undisclosedPostalstreetPercentageThirdParty{})                          \\
                          &                      & City               & \undisclosedPostalcityCounts{}               & \undisclosedPostalcityTLS{} (\undisclosedPostalcityPercentageTLS{})                             & \undisclosedPostalcityFirstParty{} (\undisclosedPostalcityPercentageFirstParty{})                             & \undisclosedPostalcityThirdParty{} (\undisclosedPostalcityPercentageThirdParty{})                              \\
                          &                      & County             & \undisclosedPostalcountyCounts{}             & \undisclosedPostalcountyTLS{} (\undisclosedPostalcountyPercentageTLS{})                         & \undisclosedPostalcountyFirstParty{} (\undisclosedPostalcountyPercentageFirstParty{})                         & \undisclosedPostalcountyThirdParty{} (\undisclosedPostalcountyPercentageThirdParty{})                          \\
                          &                      & ZIP Code           & \undisclosedPostalzipCounts{}                & \undisclosedPostalzipTLS{} (\undisclosedPostalzipPercentageTLS{})                               & \undisclosedPostalzipFirstParty{} (\undisclosedPostalzipPercentageFirstParty{})                               & \undisclosedPostalzipThirdParty{} (\undisclosedPostalzipPercentageThirdParty{})                                \\ 
\hline
Protected Classifications &                      & Gender             & \undisclosedGenderCounts{}                   & \undisclosedGenderTLS{} (\undisclosedGenderPercentageTLS{})                                     & \undisclosedGenderFirstParty{} (\undisclosedGenderPercentageFirstParty{})                                     & \undisclosedGenderThirdParty{} (\undisclosedGenderPercentageThirdParty{})                                      \\
                          &                      & Date of Birth      & \undisclosedDateofbirthCounts{}              & \undisclosedDateofbirthTLS{} (\undisclosedDateofbirthPercentageTLS{})                           & \undisclosedDateofbirthFirstParty{} (\undisclosedDateofbirthPercentageFirstParty{})                           & \undisclosedDateofbirthThirdParty{} (\undisclosedDateofbirthPercentageThirdParty{})                            \\ 
\hline
Geolocation               & Precise              & GPS Coordinates    & \undisclosedPreciselongitudelatitudeCounts{} & \undisclosedPreciselongitudelatitudeTLS{} (\undisclosedPreciselongitudelatitudePercentageTLS{}) & \undisclosedPreciselongitudelatitudeFirstParty{} (\undisclosedPreciselongitudelatitudePercentageFirstParty{}) & \undisclosedPreciselongitudelatitudeThirdParty{} (\undisclosedPreciselongitudelatitudePercentageThirdParty{})  \\
                          & Coarse               & GPS Coordinates    & \undisclosedCoarselongitudelatitudeCounts{}  & \undisclosedCoarselongitudelatitudeTLS{} (\undisclosedCoarselongitudelatitudePercentageTLS{})   & \undisclosedCoarselongitudelatitudeFirstParty{} (\undisclosedCoarselongitudelatitudePercentageFirstParty{})   & \undisclosedCoarselongitudelatitudeThirdParty{} (\undisclosedCoarselongitudelatitudePercentageThirdParty{})    \\
                          &                      & City               & \undisclosedCurrentcityCounts{}              & \undisclosedCurrentcityTLS{} (\undisclosedCurrentcityPercentageTLS{})                           & \undisclosedCurrentcityFirstParty{} (\undisclosedCurrentcityPercentageFirstParty{})                           & \undisclosedCurrentcityThirdParty{} (\undisclosedCurrentcityPercentageThirdParty{})                            \\
                          &                      & County             & \undisclosedCurrentcountyCounts{}            & \undisclosedCurrentcountyTLS{} (\undisclosedCurrentcountyPercentageTLS{})                       & \undisclosedCurrentcountyFirstParty{} (\undisclosedCurrentcountyPercentageFirstParty{})                       & \undisclosedCurrentcountyThirdParty{} (\undisclosedCurrentcountyPercentageThirdParty{})                        \\
                          &                      & ZIP Code           & \undisclosedCurrentzipCounts{}               & \undisclosedCurrentzipTLS{} (\undisclosedCurrentzipPercentageTLS{})                             & \undisclosedCurrentzipFirstParty{} (\undisclosedCurrentzipPercentageFirstParty{})                             & \undisclosedCurrentzipThirdParty{} (\undisclosedCurrentzipPercentageThirdParty{})                              \\ 
\hline
Professional              &                      & Job                & \undisclosedJobCounts{}                      & \undisclosedJobTLS{} (\undisclosedJobPercentageTLS{})                                           & \undisclosedJobFirstParty{} (\undisclosedJobPercentageFirstParty{})                                           & \undisclosedJobThirdParty{} (\undisclosedJobPercentageThirdParty{})                                            \\
                          &                      & Company            & \undisclosedCompanyCounts{}                  & \undisclosedCompanyTLS{} (\undisclosedCompanyPercentageTLS{})                                   & \undisclosedCompanyFirstParty{} (\undisclosedCompanyPercentageFirstParty{})                                   & \undisclosedCompanyThirdParty{} (\undisclosedCompanyPercentageThirdParty{})                                    \\ 
\hline
Education                 &                      & University         & \undisclosedCollegeCounts{}                  & \undisclosedCollegeTLS{} (\undisclosedCollegePercentageTLS{})                                   & \undisclosedCollegeFirstParty{} (\undisclosedCollegePercentageFirstParty{})                                   & \undisclosedCollegeThirdParty{} (\undisclosedCollegePercentageThirdParty{})                                    \\
Sensory Data              &                      & Sensor Readings    & \undisclosedSensordataCounts{}               & \undisclosedSensordataTLS{} (\undisclosedSensordataPercentageTLS{})                             & \undisclosedSensordataFirstParty{} (\undisclosedSensordataPercentageFirstParty{})                             & \undisclosedSensordataThirdParty{} (\undisclosedSensordataPercentageThirdParty{})                              \\
\hline
\end{tabular}

    \begin{tablenotes}
      \small
      \item `\# Apps' denotes the total number of apps that did not disclose the specific PII out of a total of 80 apps that provided valid responses to VCRs. Percentages denote the proportion out of the total number of apps that did not disclose the specific PII.
    \end{tablenotes}

  \end{threeparttable}
\end{table*}

In terms of the specific pieces of personal information, we observed the collection, but not the disclosure, of device-specific identifiers, such as the Android Advertising ID (AAID), by \undisclosedProvidedSpecificDeviceIdentifiersCount~apps, app-specific identifiers, such as the Android ID, by \undisclosedProvidedSpecificAppIdentifiersCount~apps, coarse GPS coordinates (i.e., with a granularity up to a certain neighborhood) by \undisclosedProvidedSpecificCoarseLatLonCount, ZIP code by \undisclosedProvidedSpecificZIPCount, the name of the city by \undisclosedProvidedSpecificCityCount~apps, precise GPS coordinates (i.e., that point to a specific building) by \undisclosedProvidedSpecificPreciseLatLonCount, parts of postal address by \undisclosedProvidedSpecificResidenceRecordsCount, user's phone number by \undisclosedProvidedSpecificPhoneCount, information about a user's contacts by \undisclosedProvidedSpecificContactRecordsCount~apps, and so on. 

We examined the network transmission logs for the \ResponseWithoutDataCount~apps developed by companies that told us that they did not hold any data on us; only one appeared to not actually collect any data. The remaining \undisclosedClaimedNotCollectedCount~collected data across a range of CCPA-defined categories of personal information, in particular, identifiers (\undisclosedClaimedNotCollectedIdentifiersCount), geolocation (\undisclosedClaimedNotCollectedGeolocationCount), and sensory data (\undisclosedClaimedNotCollectedSensoryCount). More specifically, all \undisclosedClaimedNotCollectedAAIDCount~apps collected the AAID, \undisclosedClaimedNotCollectedIPCount~collected our IP address, and one collected a device-identifying ID generated by the Branch.io SDK. Furthermore, one of the apps collected, but did not disclose the collection of precise GPS coordinates, and \undisclosedClaimedNotCollectedCoarseGeoCount~apps collected coarse geolocation data that pinpointed the specific city, neighborhood, or ZIP code, where the device was physically located. Finally, \undisclosedClaimedNotCollectedSpecificSensorCount~apps collected readings generated by the device's accelerometer, gyroscope, or magnetometer sensors.

\begin{figure}[t]
  \centering
  \includegraphics[width=\linewidth]{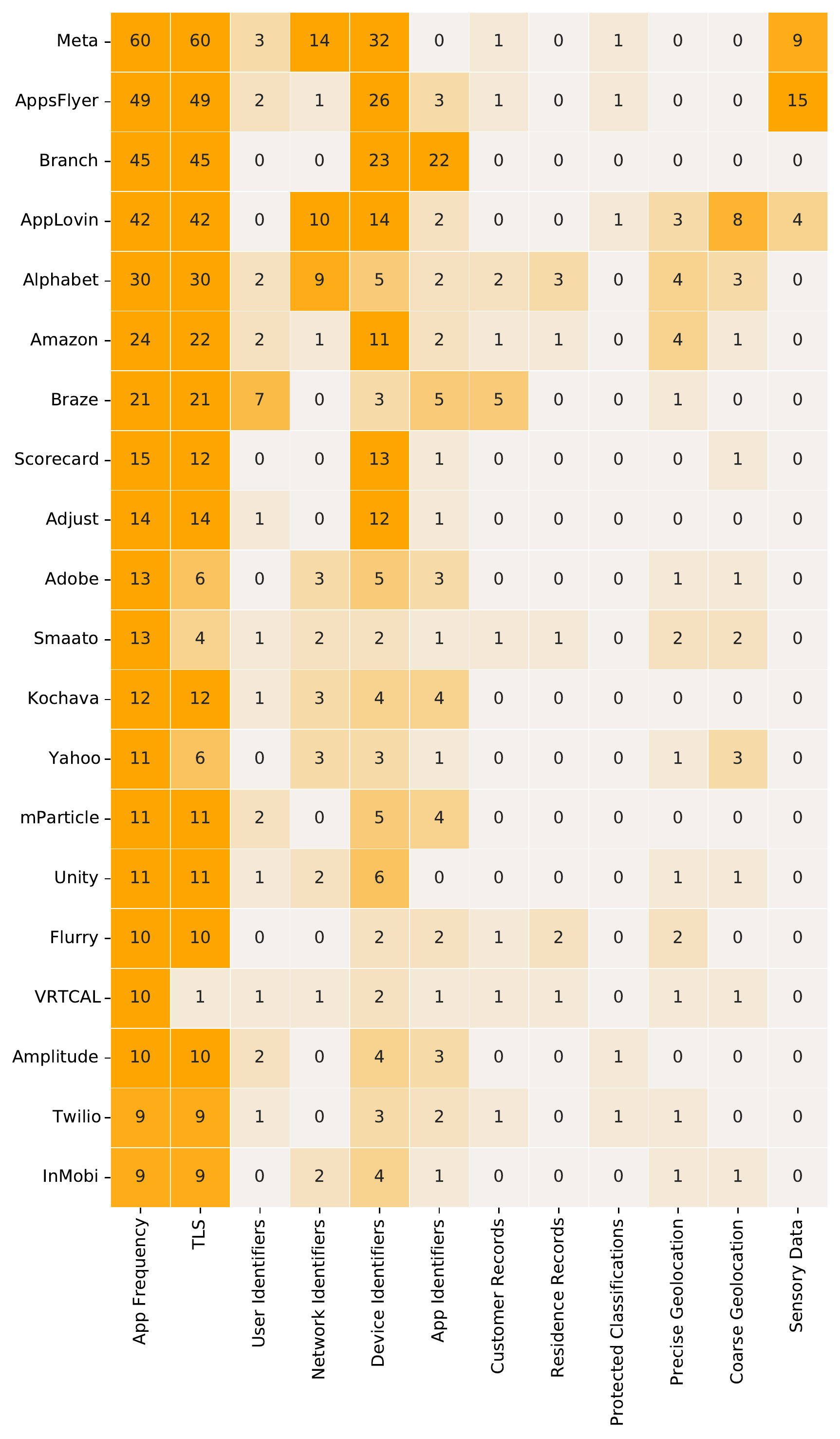}
  \caption{Top 20 Third-Party Data Recipients. \textmd{\small Each number represents the unique number of apps, from which the entity collected a specific category of personal data. The first two columns display the number of unique apps sharing any category of PII, and whether it was via TLS.}}
  \label{fig:entities}
\end{figure}

Table~\ref{tab:undisclosedPII} summarizes the undisclosed data collection that we observed across the \ResponseYesCount~apps, for which we received a response, including information about the usage of TLS encryption, as well as the number of apps that do not disclose the categories of personal information shared with the first-party and third-party domains. We note that our results provide a lower bound on the number of pieces of collected-but-undisclosed personal information, as additional personal information collected by the apps might not have been detected during our analysis of the apps' network traffic. We additionally present the top 20 third-party recipients of personal information, as well as the number of apps that shared different categories of personal information with these entities in Figure~\ref{fig:entities}.

\subsection{Privacy Policies}

Finally, we analyzed the disclosures made in the privacy policies of tested apps. For each of the \totalCCPAAppCount{} privacy policies containing CCPA-specific information, multiple researchers from our team independently indicated which categories of personal information were collected or disclosed by each developer and to which category of recipients. Table~\ref{tab:policyAnalysis} summarizes the number of policies disclosing the collection and sharing of categories of personal information, the categories of recipients, and the inter-rater reliability scores.



\begin{table}

\small
\centering
\caption{Categories Disclosed in Privacy Policies}
\label{tab:policyAnalysis}
\resizebox{\linewidth}{!}{%

\begin{threeparttable}
\begin{tabular}{l|l|l|l|l}
\multicolumn{1}{c|}{\textbf{Prompt}}                                                                                                                       & \textbf{Category}                                                      & \textbf{Yes \#} & \textbf{No \#} & $\alpha$  \\ 
\hline
\multirow{11}{*}{\begin{tabular}[c]{@{}l@{}}Does the privacy \\policy state that \\the app developer \\collects....\end{tabular}}                          & Identifiers                                                            & 109             & 0              & —         \\
                                                                                                                                                           & Customer Records                                                       & 95              & 14             & 0.517     \\
                                                                                                                                                           & Protected Classifications                                              & 63              & 46             & 0.663     \\
                                                                                                                                                           & Commercial Information                                                 & 78              & 31             & 0.596     \\
                                                                                                                                                           & Biometric Information                                                  & 12              & 97             & 0.714     \\
                                                                                                                                                           & \begin{tabular}[c]{@{}l@{}}Network Activity\end{tabular} & 107             & 2              & 0.176     \\
                                                                                                                                                           & Geolocation Data                                                       & 97              & 12             & 0.616     \\
                                                                                                                                                           & Sensory Data                                                           & 63              & 46             & 0.373     \\
                                                                                                                                                           & Professional Information                                               & 46              & 63             & 0.726     \\
                                                                                                                                                           & Education Information                                                  & 15              & 94             & 0.616     \\
                                                                                                                                                           & Inferences                                                             & 62              & 47             & 0.542     \\ 
\hline
\multirow{11}{*}{\begin{tabular}[c]{@{}l@{}}Does the privacy \\policy state that \\the app developer \\discloses or \\shares....\end{tabular}}             & Identifiers                                                            & 103             & 6              & $<0$      \\
                                                                                                                                                           & Customer Records                                                       & 81              & 28             & 0.183     \\
                                                                                                                                                           & Protected Classifications                                              & 49              & 60             & 0.411     \\
                                                                                                                                                           & Commercial Information                                                 & 65              & 44             & 0.445     \\
                                                                                                                                                           & Biometric Information                                                  & 8               & 101            & 0.579     \\
                                                                                                                                                           & \begin{tabular}[c]{@{}l@{}}Network Activity\end{tabular} & 98              & 11             & 0.099     \\
                                                                                                                                                           & Geolocation Data                                                       & 84              & 25             & 0.287     \\
                                                                                                                                                           & Sensory Data                                                           & 53              & 56             & 0.275     \\
                                                                                                                                                           & Professional Information                                               & 29              & 80             & 0.625     \\
                                                                                                                                                           & Education Information                                                  & 15              & 94             & 0.605     \\
                                                                                                                                                           & Inferences                                                             & 58              & 51             & 0.434     \\ 
\hline
\multirow{13}{*}{\begin{tabular}[c]{@{}l@{}}Does the privacy \\policy state that \\the app developer \\shares personal \\information with...\end{tabular}} & Affiliates                                                             & 98              & 11             & 0.449     \\
                                                                                                                                                           & Advertising Networks                                                   & 97              & 12             & 0.356     \\
                                                                                                                                                           & Marketing                                                              & 86              & 23             & 0.517     \\
                                                                                                                                                           & Analytics                                                              & 101             & 8              & 0.275     \\
                                                                                                                                                           & Security and Fraud                                                     & 66              & 43             & 0.293     \\
                                                                                                                                                           & Payment Processors                                                     & 78              & 31             & 0.573     \\
                                                                                                                                                           & Customer Support                                                       & 55              & 54             & 0.596     \\
                                                                                                                                                           & Storage and Infrastructure                                             & 59              & 50             & 0.637     \\
                                                                                                                                                           & Search Engines                                                         & 10              & 99             & 0.347     \\
                                                                                                                                                           & Social Media                                                           & 49              & 60             & 0.599     \\
                                                                                                                                                           & Order Fulfillment                                                      & 25              & 84             & 0.559     \\
                                                                                                                                                           & Law Enforcement                                                        & 106             & 3              & 0.234     \\
                                                                                                                                                           & Unspecified Partners                                                   & 78              & 31             & 0.042    
\end{tabular}
\begin{tablenotes}
\small
\item~Column `$\alpha$' refers to Krippendorff’s alpha, a measure of inter-rater reliability.
\end{tablenotes}
\end{threeparttable}
}

\end{table}

All 109 policies disclosed the collection of identifiers and only two did not mention the collection of ``Internet activity information,'' which includes data about app interactions. Additionally, 97 (89\%) and 95 (87\%) policies disclosed the collection of geolocation data and customer records information, respectively. The broad nature of these categories entails that most developers collect and, frequently, share this information, particularly in the context of mobile apps where technical identifiers, data from sensors, and usage information can be used both to provide the required app functionality and to track users. By the same token, users do not gain much by being informed about the collection of these categories. 

We also identified the categories of personal information that the developers disclosed or sold,~\footnote{Cal. Civ. Code \S1798.140(t)(1) broadly defines `selling' as disclosing ``a consumer's personal information by the business to another business or a third party for monetary or other valuable consideration,'' i.e., even when no monetary exchange is involved.} as well as the categories of recipients of users' personal information. Although the CCPA requires companies to enumerate the recipients for each category of personal information, in practice we found that only a small number of policies did so. Therefore, we focused on locating the categories of recipients in the text of policies irrespective of which personal information they received. 

\begin{table}
\small\centering
\caption{Comparison of Observed and Disclosed Categories}
\label{tab:categories}
\begin{threeparttable}
\begin{tabular}{l|c|cc|cc}
\multicolumn{1}{c|}{\textbf{Categories}} & \textbf{Apps} & \multicolumn{2}{c|}{\textbf{Policies }} & \multicolumn{2}{c}{\textbf{VCRs }}         \\
\multicolumn{1}{c|}{}                    &                   & \rotatebox[origin=c]{-90}{\textit{Disclosed}} & \rotatebox[origin=c]{-90}{\textit{Undisclosed~}}       & \rotatebox[origin=c]{-90}{\textit{Disclosed}} & \rotatebox[origin=c]{-90}{\textit{Undisclosed~}}  \\ 
\hline
Identifiers — Collection                 & 75                & 74                 & 1                          & 22                 & 53                    \\
Identifiers — Sharing                    & 60                & 55                 & 5                          & 11                 & 49                    \\
Customer Records — Collection            & 59                & 56                 & 3                          & 12                 & 47                    \\
Customer Records — Sharing               & 16                & 13                 & 3                          & 2                  & 14                    \\
Protected Classifications — Collection   & 16                & 9                  & 7                          & 5                  & 11                    \\
Protected Classifications — Sharing      & 4                 & 2                  & 2                          & 1                  & 3                     \\
Geolocation Data — Collection            & 38                & 36                 & 2                          & 6                  & 32                    \\
Geolocation Data — Sharing               & 23                & 20                 & 3                          & 3                  & 20                    \\
Sensory Data — Collection                & 22                & 15                 & 7                          & 0                  & 22                    \\
Sensory Data — Sharing                   & 22                & 10                 & 12                         & 0                  & 22                    \\
Professional Information — Collection    & 12                & 5                  & 7                          & 3                  & 9                     \\
Professional Information — Sharing       & 1                 & 0                  & 1                          & 0                  & 1                     \\
Education Information — Collection       & 8                 & 2                  & 6                          & 1                  & 7                     \\
Education Information — Sharing          & 0                 & 0                  & 0                          & 0                  & 0                     \\
Affiliates or Subsidiaries               & 3                 & 3                  & 0                          & 0                  & 3                     \\
Advertising Networks                     & 23                & 22                 & 1                          & 5                  & 18                    \\
Marketing                                & 27                & 17                 & 10                         & 3                  & 24                    \\
Analytics                                & 49                & 46                 & 3                          & 7                  & 42                    \\
Security and Fraud                       & 3                 & 3                  & 0                          & 0                  & 3                     \\
Payment Processors                       & 2                 & 2                  & 0                          & 0                  & 2                     \\
Customer Support                         & 1                 & 0                  & 1                          & 0                  & 1                     \\
Storage and Infrastructure               & 26                & 15                 & 11                         & 2                  & 24                    \\
Search Engines                           & 5                 & 0                  & 5                          & 0                  & 5                     \\
Social Media                             & 35                & 15                 & 20                         & 1                  & 34                   
\end{tabular}
\begin{tablenotes}
\small\item~`Apps' denotes the number of apps observed collecting or sharing a specific category of PII, or the number of apps that transmitted some PII to a specific third-party recipient, while `Disclosed' indicates how many of these disclosed that collection or sharing in a privacy policy or a VCR.
\end{tablenotes}
\end{threeparttable}
\end{table}

Unsurprisingly, we found that the most frequently collected categories of personal information are also the most frequently shared. In particular, 103 (94\%) policies disclosed the sharing of identifiers, 98 (90\%) disclosed the sharing of internet activity information, and 84 (77\%) disclosed the sharing of geolocation data. With respect to recipients, almost every privacy policy (106 or 97\%) stated that the company might share users' personal information with law enforcement, if legally compelled. We also observed analytics providers (93\% of policies), advertising networks (89\%), and marketing partners (79\%) being disclosed as the stated recipients of personal information from the apps' users. 

For many categories, we did not attain a significant level of inter-rater reliability (Krippendorff's $\alpha$ in Table~\ref{tab:policyAnalysis}). We attribute this result to the broad nature of some categories. For instance, there is a significant overlap between the `identifiers' and `customer records' categories. Recipients of personal information also commonly fall into similar categories, e.g., many companies that provide advertising also offer analytics and marketing solutions. Finally, although some companies used the CCPA-defined categories of personal information to describe their data collection and sharing practices, others relied on their own categorizations, and the CCPA does not define the categories of third-party recipients, further decreasing the consistency between policies written by different developers.

We observed the highest inter-rater agreement regarding the collection of professional or employment-related data (Krippendorff's $\alpha$ =  0.726), biometric data (0.714), and protected classifications (0.663). In general, a Krippendorff's alpha of .667 or higher is considered acceptable for drawing tentative conclusions~\cite{krippendorff2011computing}.

\paragraph{\textbf{Categories Comparison.}}
Finally, we compared the categories of personal information that we observed being collected and the categories of recipients with the categories disclosed by the developer in the VCR response and with the categories that we obtained from analyzing the privacy policies. We present the results of this comparison for the 80 apps that completed the VCR in Table~\ref{tab:categories}. Compared to the VCR responses, 25 (31\%) privacy policies failed to fully inform us about all of the categories of collected personal information, while only 17 (21\%) did not fully disclose the sharing of information to third parties.  


\section{Discussion}
\label{sec:disc}
Our results present several important implications for developers and policy makers with respect to the process of submitting verifiable consumer requests and ensuring accurate responses. We highlight the following areas for improvement: determining CCPA applicability, the security of consumers' personal information, and the usability, completeness, and accuracy of developers' responses. 

\subsection{Determining CCPA Applicability}
Only 71\% of selected apps included CCPA-specific disclosures in their privacy policies. As a compromise between evaluating the compliance of popular apps without burdening smaller developers that do not have to comply, we decided only to submit VCRs to those who provided CCPA-specific information in their privacy documents. However, this naturally limited the scope of our analysis and also prompted us to consider how ordinary consumers could determine which companies are covered by CCPA requirements.

We imagine that the only organizations that consumers could realistically determine to conform to the CCPA's definition of a ``business'' (see Section 2.1) are public companies that disclose revenues in earnings reports. However, this severely limits the ability of consumers to determine whether a company has to comply with the CCPA; even if everyone could easily read earnings reports, fewer than 0.01\% of companies in the U.S. are publicly traded~\cite{public-company}. Companies with a large online presence can surpass the data collection threshold if, for instance, they use cookies, other tracking technologies, or even simply record technical information from users' devices, such as IP addresses, but there is no way for consumers to know when the threshold is met. This could be addressed by requiring all companies doing business in California to state in their privacy policies whether they are subject to the CCPA.

\subsection{Authentication and Security}
Our analysis also demonstrated that many app developers did not use any identity verification mechanism beyond a proof of access to an email account; other companies required copies of government-issued identity documents and signed affidavits. Given different domains and company sizes, it is unlikely that a one-size-fit-all authentication approach will work for all organizations. However, we highlight several issues that we encountered and propose solutions. 

For apps that maintain user accounts, we suggest relying on existing authentication mechanisms to submit requests and access the provided data. At the very least, these companies should require a password to perform these actions. Ideally, these companies would also require a second authentication factor, such as a mobile push notification or a one-time password (OTP). App developers should also notify users about VCR submissions using established communication channels to help detect fraudulent requests.

Authentication is more difficult for developers that do not require the creation of user accounts to access their apps. These companies should request at least three (and possibly more) non-trivial pieces of user-specific information to match against the data already held. In the case of mobile apps, the developer could require the user to send the VCR via the app, such that the request also contains device-specific information alongside the requested user-specific information. However, developers should also provide an option to submit VCRs via other means, as a user might have already uninstalled the app or changed their device. If the company does not hold sufficient information to verify the consumer to a reasonable degree, then they should rightfully reject the request to avoid leaking consumers' personal information to unauthorized parties. Companies should also \textbf{not} request copies of government-issued IDs for authentication, as most organizations would not (and, ideally, should not) have access to unique ID numbers to match against; information in photos, such as name or birthdate, can be easily digitally altered.

Finally, once the developer successfully confirms the identity of the consumer, they should take necessary precautions to secure access to and transmission of consumer's personal information. In addition to existing authentication mechanisms and, ideally, two-factor authentication, developers should employ TLS, use download links with a time expiration, and secure files using a password set by the consumer beforehand. Although none of these measures can fully prevent the leakage of personal information, they can definitely increase the cost for attackers attempting to fraudulently gain access to consumers' sensitive information. 


\subsection{Usability, Completeness, and Accuracy}
We also discovered that VCR responses from app developers noticeably varied in their format and contents. For instance, although 97\% of companies that completed our requests provided specific pieces of personal information, that proportion dropped to 35\% for categories of third parties. Furthermore, only 7 companies provided a choice to receive the data either in a human-readable (e.g., TXT) or a machine-readable format (e.g., JSON).

We believe that regulators should issue more guidance to businesses when it comes to the logistics of providing personal information back to consumers. Besides questions of authentication and security, regulators should provide examples of categorizations that developers could use in responding to VCRs. For instance, although the text of the CCPA mentions covered categories of personal information, similar categories for third parties or sources of collection are absent. Many businesses use CCPA-defined categories of personal information in their policies and VCR responses and, thus, similar taxonomies would be beneficial in other contexts. We believe that to achieve greater transparency, the CCPA should also require companies to disclose names of third parties with whom they share personal information, as opposed to only requiring the categories to be disclosed. 

With respect to the accuracy of responses containing specific pieces of personal information, we discovered that developers would often collect but not disclose identifiers, geolocation data, and sensory data. As is already the case in newer versions of Android, developers should not be allowed to collect persistent non-resettable identifiers from consumers' phones, such as hardware identifiers. Instead, developers and third-party libraries should only gain access to dedicated, resettable identifiers, specifically, the Android Advertising ID (AAID). Regulators should also remind developers that device identifiers, even resettable ones, constitute personal information under the CCPA and, therefore, have to be disclosed upon receipt of a verifiable consumer request. Developers should also be reminded that the collection of such identifiers increases their chance of becoming subject to the CCPA once they reach the predefined data collection threshold. Providing more examples to developers, especially in the context of mobile apps, could help clarify what information and at which level of granularity constitutes personal information under the CCPA.

Finally, the CCPA's ``right to know'' encompasses two distinct privacy rights: the right of access and the right to data portability. Although both rights can provide access to personal information held by a business, they serve different purposes. Whereas data provided under the right to data portability should be easily imported or transmitted to another service, data provided under the right of access should be comprehensible to the consumer to whom the data pertains. As these two privacy rights are not differentiated under the CCPA the same way they are, for instance, under the GDPR, businesses provide responses mainly in the machine-readable formats that are easier to export, such as JSON. However, such formats are unlikely to be easily usable by ordinary consumers. We therefore argue that the CCPA could be enhanced by differentiating between the two rights and by providing guidelines to developers about the best practices and formats to use when responding to requests under each of these rights.


\section{Limitations}
\label{sec:limits}
We investigated the extent to which Android app developers comply with the provisions of the CCPA that require them to disclose their data sharing practices in privacy policies and in response to consumers' access requests. As our objective was to select apps that we reasonably inferred to fall under the CCPA definition of a ``business,'' it is important to note that the resulting sample of apps is not meant to be representative. Our results, therefore, do not generalize to the entire population of Android apps and do not necessarily provide insights about the data collection and sharing behaviors of other apps. 

As we previously explained in Section~\ref{sec:methods}, we tested the apps and interacted with developers using pseudonyms. We acknowledge that some companies may not have automated systems to process CCPA-related requests, and therefore processing our VCRs may have imposed costs on the employees responding to requests. However, as in related studies~\cite{ausloos2018shattering, herrmann2016obtaining, norris2017exercising, kroger2020app, urban2019study, wong2019right}, we believe that our approach was necessary to investigate the quality of the VCR responses under realistic conditions and to mitigate research participation effects~\cite{mccambridge2014systematic}. Furthermore, we believe that that business interests in this regard are outweighed by the public interest in understanding the effectiveness of CCPA rights and raising awareness around existing issues.

Finally, the developments in privacy regulation will necessitate further work in understanding how changes in specific scopes and provisions translate into differences in compliance of different businesses. In particular, most of the provisions of the California Privacy Rights Act (CPRA) revising the CCPA will become operative on January 1, 2023, with enforcement commencing on July 1, 2023. We believe that future work should continue examining the application of and compliance with the new privacy regimes to guide the development of further consumer data protection laws. 

\begin{acks}
This work was supported by the U.S. National Science Foundation (under grant CNS-1817248), the National Security Agency (under contract H98230-18-D-0006), the Center for Long-Term Cybersecurity (CLTC) at U.C. Berkeley, and by CITRIS and the Banatao Institute at the University of California. We would like to thank Brandie Nonnecke and Liam Webster for feedback, as well as Refjoh\"{u}rs Lykkewe.
\end{acks}

\bibliographystyle{ACM-Reference-Format}
\bibliography{references}

\appendix
\section{VCR Email Templates}
\label{sec:vcr_emails}
This appendix contains the email templates that we used to submit verifiable consumer requests to app developers, as well as the conditions under which we sent it. Note that we cited the provision \textit{Cal. Civil Code 1798.140} in the template emails to direct the developers to the list of categories predefined by the CCPA to facilitate their response and to improve the consistency of categorization across different companies. 


\subsection{Initial Request}
Email template used to initiate the VCR.\\

\small\noindent
Dear Privacy Compliance Officer,

My name is [\textit{name}]. I live in California and I am exercising my data access rights under the California Consumer Privacy Act (CCPA) to obtain a copy of the categories and the specific pieces of personal information that [\textit{company}] has collected about me.

I’m requesting a copy of any and all of the records you have pertaining to me including (but not limited to):
\begin{enumerate}
\item Specific pieces of personal information and any persistent identifiers that you have collected about me including all information or content provided or posted by me, any information you have collected about me, or any personal information you have obtained or acquired about me from a third party business or service provider;
\item Categories of personal information you have collected about me pursuant to the enumerated list of categories in Cal. Civil Code 1798.140(o);
\item Categories of sources from which my personal information is collected;
\item Categories of personal information that you have sold or disclosed for a business purpose about me by each category of personal information enumerated in Cal. Civil Code 1798.140(o);
\item Third parties to whom my personal information was sold or disclosed for a business purpose; and
\item The business or commercial purpose for collecting or selling my personal information.
\end{enumerate}

I expect a confirmation of receipt within 10 business days and information about how [\textit{company}] will process my request, sent to this email address. Please let me know if you need any more information from me as soon as possible. 

If you believe that you are not subject to the CCPA, please reply back as soon as possible and let me know why you believe the CCPA does not apply in this case.

\noindent
Sincerely,

\noindent
[\textit{Name}]
\normalsize

\subsection{Unable to Perform Request}
Company has directed us to use an alternative method to submit VCR that does not provide access to the full records.\\

\small\noindent
Dear [\textit{Name of Privacy Compliance Officer}],

Thank you for your reply. Unfortunately, the [\textit{alternative request method}] that you have directed me to use to submit my request does not allow me to fully exercise my data access rights under the California Consumer Privacy Act.

Specifically, the [\textit{alternative request method}] does not allow me to request a copy of the following records you have pertaining to me:

\textit{(Select and include the appropriate ones in the email)}

\begin{enumerate}
\item Specific pieces of personal information and any persistent identifiers that you have collected about me including all information or content provided or posted by me, any information you have collected about me, or any personal information you have obtained or acquired about me from a third party business or service provider;
\item Categories of personal information you have collected about me pursuant to the enumerated list of categories in Cal. Civil Code 1798.140(o);
\item Categories of sources from which my personal information is collected;
\item Categories of personal information that you have sold or disclosed for a business purpose about me by each category of personal information enumerated in Cal. Civil Code 1798.140(o);
\item Third parties to whom my personal information was sold or disclosed for a business purpose; and
\item The business or commercial purpose for collecting or selling my personal information.
\end{enumerate}

Please let me know how I should proceed as soon as possible.

\noindent
Sincerely,

\noindent
[\textit{Name}]
\normalsize

\subsection{Missing Information Request}
Company responded to our VCR without providing all of the requested information.\\

\small\noindent
Dear [\textit{Name of Privacy Compliance Officer}],

Thank you for your reply. Unfortunately, the copy of the records that I have received does not contain all of the requested information. Specifically, I have not received a copy of the following records you have pertaining to me:

\textit{(Select and include the appropriate ones in the email)}

\begin{enumerate}
\item Specific pieces of personal information and any persistent identifiers that you have collected about me including all information or content provided or posted by me, any information you have collected about me, or any personal information you have obtained or acquired about me from a third party business or service provider;
\item Categories of personal information you have collected about me pursuant to the enumerated list of categories in Cal. Civil Code 1798.140(o);
\item Categories of sources from which my personal information is collected;
\item Categories of personal information that you have sold or disclosed for a business purpose about me by each category of personal information enumerated in Cal. Civil Code 1798.140(o);
\item Third parties to whom my personal information was sold or disclosed for a business purpose; and
\item The business or commercial purpose for collecting or selling my personal information.
\end{enumerate}

Please let me know how I should proceed as soon as possible.

\noindent
Sincerely,

\noindent
[\textit{Name}]
\normalsize

\subsection{Account Holder Verification Request}
We created an account with the app and the developer required us to furnish documentation to verify our identity that we could not provide.\\

\small\noindent
Dear [\textit{Name of Privacy Compliance Officer}],

Thank you for your reply. Unfortunately, I prefer not to provide the information that you have requested to verify my identity, as I believe it to be invasive and beyond the requirements of the CCPA.

As an account holder with [\textit{company}], I would prefer verifying my identity using existing authentication practices for my account per CCR § 999.324(a). For your convenience, the [\textit{email address OR username}] associated with my account is [\textit{email address OR username}].

Please let me know if you need any more information from me as soon as possible.

\noindent
Sincerely,

\noindent
[\textit{Name}]
\normalsize

\subsection{Account Non-Holder Verification Request}
We \textit{did not} create an account with the app and the developer required us to furnish documentation to verify our identity that we could not provide.\\

\small\noindent
Dear [\textit{Name of Privacy Compliance Officer}],

Thank you for your reply. Unfortunately, I prefer not to provide the information that you have requested to verify my identity, as I believe it to be invasive and beyond the requirements of the CCPA. 

Instead, I would prefer verifying my identity by matching the following three pieces of personally identifiable information that I have previously provided to [\textit{company}] per CCR § 999.325(b) and (c): 

\textit{(Select and include the appropriate ones in the email)}

\begin{enumerate}
\item  \textit{PII1 Type: PII1 Value}
\item  \textit{PII2 Type: PII2 Value}
\item  \textit{PII3 Type: PII3 Value}
\end{enumerate}

Please let me know if you need any more information from me as soon as possible.

\noindent
Sincerely,

\noindent
[\textit{Name}]
\normalsize

\subsection{First Follow-Up}
Company did not respond to our initial request in 10 business days.\\

\small\noindent
Dear Privacy Compliance Officer,

My name is [\textit{name}] and I am following up on a request I made on [\textit{date}] to access the personal information that [\textit{company}] has collected about me. I was expecting to receive a confirmation of receipt and information about how [\textit{company}] would process my request within 10 business days per 11 CCR § 999.313(a). 
For your convenience, my original request is as follows:

\noindent
I’m requesting a copy of any and all of the records you have pertaining to me including (but not limited to):
\begin{enumerate}
\item Specific pieces of personal information and any persistent identifiers that you have collected about me including all information or content provided or posted by me, any information you have collected about me, or any personal information you have obtained or acquired about me from a third party business or service provider;
\item Categories of personal information you have collected about me pursuant to the enumerated list of categories in Cal. Civil Code 1798.140(o);
\item Categories of sources from which my personal information is collected;
\item Categories of personal information that you have sold or disclosed for a business purpose about me by each category of personal information enumerated in Cal. Civil Code 1798.140(o);
\item Third parties to whom my personal information was sold or disclosed for a business purpose; and
\item The business or commercial purpose for collecting or selling my personal information.
\end{enumerate}
\noindent

I expect a reply to this email address as soon as possible. If you believe that you are not subject to the California Consumer Privacy Act (CCPA), please reply back as soon as possible and let me know why you believe the CCPA does not apply in this case.

\noindent
Sincerely,

\noindent
[\textit{Name}]
\normalsize

\subsection{Second Follow-Up}
Company did not respond to our first follow-up email in 10 business days.

\small\noindent
Dear Privacy Compliance Officer,

My name is [\textit{name}] and I am following up on a request I originally made on [\textit{date}] to access the personal information that [\textit{company}] has collected about me. I have previously followed up about my request on [\textit{date}], but I have not heard back from you. I was expecting to receive a confirmation of receipt and information about how [\textit{company}] would process my request within 10 business days per 11 CCR § 999.313(a). My original request is as follows:\\
\noindent
I’m requesting a copy of any and all of the records you have pertaining to me including (but not limited to):
\begin{enumerate}
\item Specific pieces of personal information and any persistent identifiers that you have collected about me including all information or content provided or posted by me, any information you have collected about me, or any personal information you have obtained or acquired about me from a third party business or service provider;
\item Categories of personal information you have collected about me pursuant to the enumerated list of categories in Cal. Civil Code 1798.140(o);
\item Categories of sources from which my personal information is collected;
\item Categories of personal information that you have sold or disclosed for a business purpose about me by each category of personal information enumerated in Cal. Civil Code 1798.140(o);
\item Third parties to whom my personal information was sold or disclosed for a business purpose; and
\item The business or commercial purpose for collecting or selling my personal information.
\end{enumerate}
\noindent

I expect a reply to this email address as soon as possible.  If you believe that you are not subject to the California Consumer Privacy Act (CCPA), please reply back as soon as possible and let me know why you believe the CCPA does not apply in this case.

\noindent
Sincerely,

\noindent
[\textit{Name}]
\normalsize
\section{Codebook}
\label{sec:codebook}
Tables~\ref{tab:piiCodes} and~\ref{tab:thirdCodes} include the codebook that we used to perform a qualitative analysis of disclosures in privacy policies. We use the categories of personal information defined in Cal. Civil Code 1798.140 to represent the codes for the collection and sharing in Table~\ref{tab:piiCodes}. Table~\ref{tab:thirdCodes} contains our codes for the categories of third parties.

For each privacy policy, coders saw the following prompts:

\begin{itemize}
  \item Does this app developer include disclosures that reference the CCPA, either as part of the general privacy policy or as a standalone document?
  \item Does the privacy policy state that the app developer collects [\textit{PII Code}]?
  \item Does the privacy policy state that the app developer discloses or shares [\textit{PII Code}]?
  \item Does the privacy policy state that the app developer shares personal information with [\textit{Third Party Code}]?
\end{itemize}

\begin{table}[t]
\small\centering
\caption{Personally-identifiable Information (PII) Codes}
\label{tab:piiCodes}
\begin{tabular}{l|l}
\textbf{PII Code}                                                                                                                              & \textbf{Description}                                                                                                                                                                                                                                                                                                                                                                                                                                    \\ 
\hline
\textbf{Identifiers}                                                                                                                           & \begin{tabular}[c]{@{}l@{}}Real name, alias, postal address, unique~perso-\\nal identifier, online~identifier, IP address,~email \\address, account name,~social security number, \\driver’s license number,~passport~number,~or \\other similar identifiers.\end{tabular}                                                                                                                                                                              \\ 
\hline
\begin{tabular}[c]{@{}l@{}}\textbf{Customer }\\\textbf{Records}\end{tabular}                                                                   & \begin{tabular}[c]{@{}l@{}}Name, signature, social security number,~physical~\\characteristics or description, address,~telephone \\number,~passport number, driver’s license or state\\identification card number,~insurance policy~num-\\ber, education,~employment, employment history, \\bank account number, credit card number, debit \\card number, or any other financial information, \\medical or health insurance~information.\end{tabular}  \\ 
\hline
\begin{tabular}[c]{@{}l@{}}Characteristics of \\\textbf{Protected~}\\\textbf{Classifications~}\\under~California \\or Federal Law\end{tabular} & \begin{tabular}[c]{@{}l@{}}Age, race, color, ancestry, national origin,~citizen-\\ship, religion or creed,~marital status,~medical~\\condition, physical or mental~disability, sex,~gen-\\der, gender identity, gender expression, pregnan-\\cy or childbirth and related medical conditions, \\sexual orientation,~veteran or military status,~\\genetic information (including familial genetic \\information).\end{tabular}                          \\ 
\hline
\begin{tabular}[c]{@{}l@{}}\textbf{Commercial }\\\textbf{Information}\end{tabular}                                                             & \begin{tabular}[c]{@{}l@{}}Records of personal property, products or servi-\\ces purchased, obtained, or considered, or other\\purchasing or consuming histories or~tendencies.\end{tabular}                                                                                                                                                                                                                                                            \\ 
\hline
\begin{tabular}[c]{@{}l@{}}\textbf{Biometric}\\\textbf{Information}\end{tabular}                                                               & \begin{tabular}[c]{@{}l@{}}Genetic, physiological, behavioral, and biological \\characteristics, or activity patterns used to extract \\a template or other identifier or identifying infor-\\mation,~such as, fingerprints, faceprints, and~\\voiceprints, iris or retina scans, keystroke, gait, \\or other physical patterns,~sleep, health, or~exer-\\cise data.\end{tabular}                                                                       \\ 
\hline
\begin{tabular}[c]{@{}l@{}}\textbf{Network }\\\textbf{Activity }\\\end{tabular}                                                                & \begin{tabular}[c]{@{}l@{}}Browsing history, search history, or information \\regarding a consumer’s interaction with a~website,\\application, or advertisement.\end{tabular}                                                                                                                                                                                                                                                                           \\ 
\hline
\begin{tabular}[c]{@{}l@{}}\textbf{Geolocation}\\\textbf{Data}\end{tabular}                                                                    & \begin{tabular}[c]{@{}l@{}}Information such as physical location or~move-\\ments.\end{tabular}                                                                                                                                                                                                                                                                                                                                                          \\ 
\hline
\begin{tabular}[c]{@{}l@{}}\textbf{Sensory}\\\textbf{Data}\end{tabular}                                                                        & \begin{tabular}[c]{@{}l@{}}Audio, electronic, visual, thermal, olfactory, or~\\similar information.\end{tabular}                                                                                                                                                                                                                                                                                                                                        \\ 
\hline
\begin{tabular}[c]{@{}l@{}}\textbf{Professional}\\\textbf{Information}\end{tabular}                                                            & \begin{tabular}[c]{@{}l@{}}Information such as current or past job history or \\performance evaluations.\\\end{tabular}                                                                                                                                                                                                                                                                                                                                 \\ 
\hline
\begin{tabular}[c]{@{}l@{}}\textbf{Education}\\\textbf{Information}\end{tabular}                                                               & \begin{tabular}[c]{@{}l@{}}Education records directly related to a student \\maintained by an educational institution or party \\acting on its behalf,~such as grades, transcripts, \\class lists,~student schedules, student identifica-\\tion codes, student financial information, or stu-\\dent disciplinary records.\end{tabular}                                                                                                                  \\ 
\hline
\textbf{Inferences}                                                                                                                            & \begin{tabular}[c]{@{}l@{}}Consumer’s~preferences, characteristics, psycho-\\logical trends,~predispositions, behavior, attitudes,~\\intelligence,~abilities,~or aptitudes.\end{tabular}                                                                                                                                                                                                                                                               
\end{tabular}
\end{table}
\begin{table}
\small\centering
\caption{Third-Party Data Recipients Codes}
\label{tab:thirdCodes}
\begin{tabular}{l|l}
\multicolumn{1}{c|}{\textbf{3\textsuperscript{rd} Party~Code}}                         & \textbf{Description}                                                                                                                                                                                                                                                      \\ 
\hline
\begin{tabular}[c]{@{}l@{}}\textbf{Affiliated }\\\textbf{Companies}\end{tabular}       & \begin{tabular}[c]{@{}l@{}}Companies related to the app developer through~\\ownership, such as when the app developer holds~\\a stake in the company (e.g., a subsidiary) or~\\when another third party controls both the com-\\pany and the app developer.\end{tabular}  \\ 
\hline
\begin{tabular}[c]{@{}l@{}}\textbf{Advertising~}\\\textbf{Networks}\end{tabular}       & \begin{tabular}[c]{@{}l@{}}Connect advertisers to websites or apps~(the \\“publishers”) that want to host advertisements.\end{tabular}                                                                                                                                    \\ 
\hline
\begin{tabular}[c]{@{}l@{}}\textbf{Marketing }\\\textbf{Providers}\end{tabular}        & \begin{tabular}[c]{@{}l@{}}Offer products, services, or other promotions to~\\the app’s users, for instance, by calling, texting or~\\emailing them with marketing messages. ~\end{tabular}                                                                               \\ 
\hline
\begin{tabular}[c]{@{}l@{}}\textbf{Analytics}\\\textbf{Providers}\end{tabular}         & \begin{tabular}[c]{@{}l@{}}Capture data about the app’s audience in order to~\\identify unique users, track their interactions, and~\\record their behavior for the purpose of improving~\\the app, informing company strategy, or general~\\research.\end{tabular}       \\ 
\hline
\begin{tabular}[c]{@{}l@{}}\textbf{Security and }\\\textbf{Fraud}\end{tabular}         & \begin{tabular}[c]{@{}l@{}}Provide tools, such as identity verification and\\fraud detection, to prevent fraudulent activity,\\improve app security, enforce terms of service,\\and protect users and property.~\end{tabular}                                             \\ 
\hline
\begin{tabular}[c]{@{}l@{}}\textbf{Payment }\\\textbf{Processors~}\end{tabular}        & \begin{tabular}[c]{@{}l@{}}Enable merchants to sell products and accept~in-\\app card payments.~\end{tabular}                                                                                                                                                             \\ 
\hline
\begin{tabular}[c]{@{}l@{}}\textbf{Customer }\\\textbf{Support}\end{tabular}           & \begin{tabular}[c]{@{}l@{}}Provide tools to collect, organize, respond to, and \\report on customer support requests tounder-\\stand user needs, provide assistance,~and \\streamline communication.\end{tabular}                                                         \\ 
\hline
\begin{tabular}[c]{@{}l@{}}\textbf{Storage and }\\\textbf{Infrastructure}\end{tabular} & \begin{tabular}[c]{@{}l@{}}Provide services, such as data hosting, cloud~\\storage, load balancing and other infrastructure~\\to optimize content delivery and performance.\end{tabular}                                                                                  \\ 
\hline
\begin{tabular}[c]{@{}l@{}}\textbf{Search }\\\textbf{Engines}\end{tabular}             & \begin{tabular}[c]{@{}l@{}}Collect, organize and enable the search for~\\content online, including information generated~\\by users interacting with the app or other users.~\end{tabular}                                                                                \\ 
\hline
\begin{tabular}[c]{@{}l@{}}\textbf{Social Media}\\\textbf{Platforms}\end{tabular}      & \begin{tabular}[c]{@{}l@{}}Provide technologies and means of communi-\\cation, through which users create and share \\information and ideas in online communities.~\end{tabular}                                                                                          \\ 
\hline
\begin{tabular}[c]{@{}l@{}}\textbf{Order }\\\textbf{Fulfillment}\end{tabular}          & Process orders and deliver products to customers. ~                                                                                                                                                                                                                       \\ 
\hline
\begin{tabular}[c]{@{}l@{}}\textbf{Law }\\\textbf{Enforcement}\end{tabular}            & \begin{tabular}[c]{@{}l@{}}Sharing to comply with a legal obligation or a~\\request from regulators, courts, law enforcement,~\\and other governmental agencies.\end{tabular}                                                                                             \\ 
\hline
\begin{tabular}[c]{@{}l@{}}\textbf{Unspecified }\\\textbf{Partners}\end{tabular}       & \begin{tabular}[c]{@{}l@{}}Sharing with unspecified partners and~\\service providers.\end{tabular}                                                                                                                                                                       
\end{tabular}
\end{table}
\section{Data Taxonomy}
\label{sec:data_taxonomy}

Table~\ref{tab:taxonomy} enumerates the 7 categories of personal information defined in the CCPA relevant to this work, our subcategories, as well as the types and values of personal information that we have predefined for each test device.


We generated pseudonymous data for \textit{User Identifiers}, \textit{Customer Records}, \textit{Protected Classifications}, \textit{Professional} and \textit{Education Information} using publicly-available random value generators, such as those found on the Random Lists~\footnote{\url{https://www.randomlists.com/}} website and the Faker~\footnote{\url{https://pypi.org/project/Faker/0.7.4/}} Python package. We obtained other types of personal information, including \textit{Device Identifiers} and \textit{Geolocation Data}, directly from our test devices. 

\begin{table*}
\small\centering
\caption{Data Taxonomy}
\label{tab:taxonomy}
\begin{threeparttable}
\begin{tabular}{l|l|l|l|l} 
\hline
\textbf{CCPA Category}                                                                & \textbf{Subcategory} & \textbf{Description}                                                                                              & \textbf{PII Types}                                                                                                            & \textbf{Example Values}                                                                                        \\ 
\hline
\textbf{Identifiers }                                                                 & User                 & Identifiers set by the user                                                                                       & Usernames, email address, website                                                                                             & schneider90christopher19                                                                                       \\
                                                                                      & Network              & \begin{tabular}[c]{@{}l@{}}Identifiers unique to \\user's network\end{tabular}                                    & IP Address, router MAC and SSID                                                                                               & 135.***.***.79, 48:**:**:**:**:06                                                                              \\
                                                                                      & Device               & \begin{tabular}[c]{@{}l@{}}Identifiers unique\\to user's device\end{tabular}                                      & \begin{tabular}[c]{@{}l@{}}Android advertising ID (AAID), \\hardware IDs, IMEI, IMSI, SIM ID, \\Wi-Fi MAC, GSFID\end{tabular} & \begin{tabular}[c]{@{}l@{}}97PAY11GN2,~359677097304580,\\58:CB:52:8B:C8:66,\\03140e43-9bb7-[...]\end{tabular}  \\
                                                                                      & App                  & \begin{tabular}[c]{@{}l@{}}Identifiers unique to \\a single app\end{tabular}                                      & \begin{tabular}[c]{@{}l@{}}Android ID, app fingerprint ID,\\identity ID\end{tabular}                                          & \begin{tabular}[c]{@{}l@{}}7892f8834ddbf2df\\1039977256339324001\end{tabular}                                  \\ 
\hline
\begin{tabular}[c]{@{}l@{}}\textbf{Customer}\\\textbf{Records }\end{tabular}          & Customer             & \begin{tabular}[c]{@{}l@{}}Information about\\the user\end{tabular}                                               & Name, phone number, height, weight                                                                                            & \begin{tabular}[c]{@{}l@{}}Christopher Schneider,\\323-448-****\end{tabular}                                   \\
                                                                                      & Contacts             & \begin{tabular}[c]{@{}l@{}}Information about \\user's contacts\end{tabular}                                       & Contact name, contact phone number                                                                                            & Scott Pratt,~415-200-****                                                                                      \\
                                                                                      & Residence            & \begin{tabular}[c]{@{}l@{}}Information about\\user's general address \\of residence\end{tabular}                  & Street, city, county, ZIP Code                                                                                                & 957 Green Causeway, Los Angeles                                                                                \\ 
\hline
\begin{tabular}[c]{@{}l@{}}\textbf{Protected }\\\textbf{Classifications}\end{tabular} & ---                  & \begin{tabular}[c]{@{}l@{}}Information protected \\under the California and \\U.S. federal laws\end{tabular}      & Gender, date of birth, age                                                                                                    & Male, 20-May-1990                                                                                              \\ 
\hline
\begin{tabular}[c]{@{}l@{}}\textbf{Geolocation}\\\textbf{Data}\end{tabular}           & Precise              & \begin{tabular}[c]{@{}l@{}}Locates a specific \\building\end{tabular}                                             & \begin{tabular}[c]{@{}l@{}}Precise longitude/latitude coordinates, \\street name\end{tabular}                                 & *****                                                                                                          \\
                                                                                      & Coarse               & \begin{tabular}[c]{@{}l@{}}Does not locate a \\specific building\end{tabular}                                     & \begin{tabular}[c]{@{}l@{}}Coarse longitude/latitude coordinates,\\city, county, ZIP Code\end{tabular}                        & *****                                                                                                          \\ 
\hline
\textbf{Sensory Data}                                                                 & ---                  & \begin{tabular}[c]{@{}l@{}}Audio, electronic, visual, \\thermal, olfactory, or \\similar information\end{tabular} & \begin{tabular}[c]{@{}l@{}}Accelerometer, gyroscope,\\magnetometer readings\end{tabular}                                      & AK0991X,~BMI160                                                                                                \\ 
\hline
\begin{tabular}[c]{@{}l@{}}\textbf{Professional}\\\textbf{Information}\end{tabular}   & ---                  & \begin{tabular}[c]{@{}l@{}}Current or past job history or \\performance evaluations\end{tabular}                  & Job, company                                                                                                                  & \begin{tabular}[c]{@{}l@{}}Clinical Psychologist,~\\Williams and Davis\end{tabular}                            \\ 
\hline
\begin{tabular}[c]{@{}l@{}}\textbf{Education}\\\textbf{Information}\end{tabular}      & ---                  & \begin{tabular}[c]{@{}l@{}}Education records directly \\related to a student\end{tabular}                         & College                                                                                                                       & Villanova University                                                                                           \\
\hline
\end{tabular}

    \begin{tablenotes}
      \small\centering
      \item Some of the values have been redacted to preserve the privacy of researchers to whom the data pertains.
    \end{tablenotes}

\end{threeparttable}
\end{table*}

\end{document}